\documentclass[11pt,letterpaper,onecolumn]{article}
\usepackage[T1]{fontenc}
\usepackage[utf8]{inputenc}
\usepackage{lmodern}
\usepackage{microtype}
\usepackage{geometry}
\usepackage{amsmath,amssymb,mathtools}
\usepackage{graphicx}
\usepackage{booktabs,longtable,array}
\usepackage{caption}
\usepackage[section]{placeins}
\usepackage{hyperref}
\hypersetup{colorlinks=true,linkcolor=blue,citecolor=blue,urlcolor=blue,
 pdfauthor={Xin Huang, Haofei Sun, Wei Han, Yunfan Yang, Huaguan Chen, Xingtao Song, and Dong Xiao},
 pdftitle={RadarFill: Observation-Constrained Gap Filling of Systematic Blind Zones in 3-D Weather Radar Reflectivity Mosaics}}
\usepackage{tabularx}
\usepackage{float}
\renewcommand{\arraystretch}{1.15}
\renewcommand{\thetable}{\Roman{table}}
\newcommand{\authorentry}[2]{%
  \begin{minipage}[t]{\linewidth}\centering
  {\normalsize\bfseries #1\par}\vspace{0.45em}
  {\footnotesize #2\par}
  \end{minipage}%
}
\begin{document}
\begin{center}
{\LARGE\bfseries RadarFill: Observation-Constrained Gap Filling of Systematic Blind Zones in 3-D Weather Radar Reflectivity Mosaics\par}
\vspace{1.4em}
\setlength{\tabcolsep}{5pt}%
\begin{tabularx}{\textwidth}{@{}XXX@{}}
\authorentry{Xin Huang}{Department of Atmospheric and Oceanic Sciences, Fudan University, Shanghai 200438, China} &
\authorentry{Haofei Sun\textsuperscript{*}}{Shanghai Typhoon Institute, China Meteorological Administration, Shanghai 200030, China} &
\authorentry{Wei Han\textsuperscript{*}}{State Key Laboratory of Severe Weather Meteorological Science and Technology (LASW), Beijing 100081, China; CMA Earth System Modeling and Prediction Center (CEMC), China Meteorological Administration, Beijing 100081, China}
\\[2.0em]
\authorentry{Yunfan Yang}{Institute of Atmospheric Physics, Chinese Academy of Sciences, Beijing 100029, China} &
\authorentry{Huaguan Chen}{Gaoling School of Artificial Intelligence, Renmin University of China, Beijing 100872, China} &
\authorentry{Xingtao Song}{School of Atmospheric Physics, Nanjing University of Information Science and Technology, Nanjing 210044, China}
\\[2.0em]
\multicolumn{3}{c}{\begin{minipage}{0.33\textwidth}\centering
{\normalsize\bfseries Dong Xiao\par}\vspace{0.45em}
{\footnotesize Key Laboratory of Cities' Mitigation and Adaptation to Climate Change in Shanghai of China Meteorological Administration (CMACC), Shanghai, China\par}
\end{minipage}}
\end{tabularx}
\end{center}
\vspace{1.0em}
\noindent\textsuperscript{*}\textbf{Corresponding authors:} Haofei Sun (\href{mailto:sunhf@typhoon.org.cn}{sunhf@typhoon.org.cn}); Wei Han (\href{mailto:hanwei@cma.gov.cn}{hanwei@cma.gov.cn}).\\
\noindent\textbf{Funding:} National Key Research and Development Program of China (2025YFE0217100).
\vspace{0.75em}

\begin{abstract}
Ground-based weather radar mosaics contain systematic blind zones, especially at low altitudes, where missing reflectivity lacks corresponding complete observations for direct verification. We present RadarFill, an observation-constrained conditional diffusion framework for gap filling in three-dimensional radar reflectivity mosaics. RadarFill conditions on partially observed reflectivity and an observation mask, enforces available observations throughout reverse sampling, and preserves valid measurements in the final field. For training and controlled evaluation, structured synthetic missingness is designed from altitude-dependent missing-data statistics and the vertical nesting of blind zones in an operational 3-D radar mosaic product. Across 4,384 held-out test times and representative convective and typhoon cases, RadarFill recovers principal precipitation structures and moderate-to-strong echoes under extensive low-level missingness and generally outperforms conventional interpolation and a learning-based reference model. For real blind-zone evaluation, RadarFill model is applied directly to Fujian Province and compared with 90 matched volume scans from an independent phased-array radar. Within the principal 1-2 km low-level blind zones, RadarFill yields lower reflectivity errors than conventional interpolation, and the gap-filled precipitation structures show broad agreement with independent observations. In two controlled nowcasting experiments, improvements in RadarFill-initialized fields persist into subsequent forecasts, with the largest benefits at low altitudes and short lead times. These results provide complementary evidence from controlled synthetic missingness, independent radar observations, and downstream nowcasting, supporting the potential of observation-constrained conditional generation for gap filling in real 3-D radar blind zones.
\end{abstract}
\vspace{0.75em}

\FloatBarrier

\section{Introduction}

WEATHER radar provides critical observations of the spatial distribution, intensity, and vertical structure of precipitation systems and is widely used for severe-weather monitoring, quantitative precipitation estimation, nowcasting, and convective-scale data assimilation \cite{ref1}, \cite{ref2}, \cite{ref3}, \cite{ref4}, \cite{ref5}, \cite{ref6}. However, ground-based weather radars cannot fully observe the three-dimensional atmosphere because of Earth curvature, beam propagation, scanning geometry, and terrain blockage \cite{ref7}, \cite{ref8}, \cite{ref9}. As the distance from a radar increases, low-elevation beams progressively rise above the surface, limiting the observation of the remote near-surface atmosphere, while beam blockage over complex terrain further enlarges low-level observation blind zones \cite{ref4}, \cite{ref7}, \cite{ref8}, \cite{ref9}, \cite{ref10}. These limitations remain even after observations from multiple radars are mosaicked. An existing assessment reported that effective radar coverage at 1-km altitude over eastern China is only approximately 27.1\% \cite{ref9}. Such systematic data gaps reduce the completeness of three-dimensional radar reflectivity products and can affect downstream applications such as quantitative precipitation estimation \cite{ref11}, \cite{ref12}. Recovering missing reflectivity within observation blind zones is therefore an important problem for improving the completeness and utility of 3-D radar mosaic products.

Early studies of missing weather radar observations relied mainly on spatial interpolation or relationships among available radar measurements. Seed and Pegram used kriging to fill gaps caused by ground clutter using surrounding valid observations \cite{ref13}, while Zhang et al. used relationships among polarimetric radar variables to correct partial beam blockage \cite{ref14}. Deep-learning methods were subsequently applied to missing-data scenarios including near-surface blind zones, temporary observation outages, and beam blockage \cite{ref11}, \cite{ref12}, \cite{ref15}, \cite{ref16}. Yin et al. exploited echo information from multiple elevation angles to recover low-elevation blocked regions \cite{ref17}. Tan and Chen used local and global echo information for generative correction of partial beam blockage \cite{ref18}, while Gong et al. trained a network with sector-shaped synthetic missingness to restore contiguous missing regions \cite{ref15}. More recently, diffusion models have also been applied to weather radar gap filling under different missing-data configurations \cite{ref19}, \cite{ref20} and to beam-blockage correction \cite{ref21}. These studies demonstrate that learning-based and generative approaches can effectively exploit available radar information for missing-data reconstruction.

For 3-D radar mosaic products, RadGF-GAN combines 3-D radar reflectivity with satellite information to recover extensive low-level missing data; its training and evaluation rely on simulated data \cite{ref22}. Rad3DGFM directly uses 3-D radar mosaics and exploits cross-altitude structural information for gap filling and cross-region evaluation \cite{ref23}. Despite these advances, systematic blind zones in operational 3-D radar mosaics present additional practical challenges. First, real blind zones exhibit pronounced vertical organization. Our analysis of CAP data shows that missingness generally decreases with altitude, while missing regions at adjacent levels exhibit strong spatial containment. Second, although direct measurements are unavailable within blind zones, valid observations in surrounding areas and at other altitudes can constrain the reconstruction and should be preserved in the completed field. Third, real blind zones generally lack corresponding complete reference fields. Existing quantitative evaluations therefore rely largely on simulated data or synthetic missingness, leaving the applicability of synthetically trained gap-filling models to real observation blind zones insufficiently validated by independent radar observations \cite{ref22}, \cite{ref23}.

Although observations are unavailable within the blind zones themselves, precipitation systems exhibit continuous three-dimensional reflectivity structures across altitude \cite{ref24}, such that observations surrounding the missing region and at other levels may retain information relevant to the missing reflectivity. For a gap-filling task, these existing observations can provide constraints for reconstruction and should remain unchanged in the completed field \cite{ref11}. Based on this consideration, we develop RadarFill for systematic blind zones in 3-D radar mosaic products and use conditional diffusion to generate the missing reflectivity. The incomplete 3-D reflectivity field and its observation mask are jointly used as conditioning information, allowing valid observations surrounding the blind zones and at other altitudes to guide reconstruction. Available observations are further imposed throughout reverse sampling and are exactly preserved in the final completed field. RadarFill is therefore designed to recover missing reflectivity within systematic blind zones rather than to regenerate the entire observed radar volume.

Accordingly, the evaluation is organized in three stages. First, structured synthetic missingness derived from the altitude-dependent missingness characteristics of CAP V3.0 and the spatial containment of missing regions between adjacent levels is used for controlled evaluation over representative weather cases and 4,384 test times. Second, the trained model is applied directly to Fujian, a region excluded from training, and evaluated within real CAP blind zones using independent phased-array radar observations that do not contribute to the CAP mosaic product. Finally, under synthetic missingness for two weather events, the completed 3-D reflectivity fields are used in Nowcast3D \cite{ref25} to assess whether improvements in the initial field persist into subsequent forecasts.

\FloatBarrier

\section{Datasets and Preprocessing}

We used two radar datasets: the CAP 3-D reflectivity mosaic product from the China Meteorological Administration Meteorological Observation Centre\textquotesingle s Weather Radar Mosaic System V3.0, used for RadarFill training and primary testing; and S-band phased-array dual-polarization weather radar observations from Minhou during Typhoon Doksuri on 28 July 2023, used for independent validation in real blind zones.

\FloatBarrier

\subsection{CAP Data}

The CAP data are obtained from the China Meteorological Administration Meteorological Observation Centre\textquotesingle s Weather Radar Mosaic System V3.0 and span January 2023 to December 2025. The product has a temporal resolution of 6 min, a horizontal resolution of 0.01$^\circ$ $\times$ 0.01$^\circ$, and 24 constant-altitude levels, spaced at 0.5 km from 0.5 to 8 km and at 1 km from 8 to 16 km. We use the domain 118.00$^\circ$E--120.56$^\circ$E and 31.00$^\circ$N--33.56$^\circ$N, corresponding to a 256 $\times$ 256 horizontal grid. Because the 0.5-km level contains substantially more missing data, RadarFill uses the 23 levels from 1 to 16 km. The mean raw missing fractions in the training samples are 2.093\%, 0.150\%, 0.020\%, and 0.0005\% at 1.0, 1.5, 2.0, and 2.5 km, respectively, with a 23-level mean of 0.099\% (Table I).

To obtain a spatially continuous observation field, originally invalid pixels are filled independently at each altitude using 2-D linear interpolation, while the original validity masks are retained for missing-data statistics. Reflectivity is clipped to 0--70 dBZ and normalized as

\begin{equation}
X_{\mathrm{norm}}=\frac{Z-40}{40}
\label{eq:1}
\end{equation}
where Z denotes radar reflectivity. \(\text{X}_{\text{norm}}\) is further clipped to the range [-1,1]. Model outputs are transformed back to physical units according to

\begin{equation}
Z=40X_{\mathrm{norm}}+40
\label{eq:2}
\end{equation}
and clipped to 0--70 dBZ. Because interpolation affects only a small fraction of voxels, the resulting field is referred to as the observation field (Observation) and is used for training and evaluation under synthetic missingness. It is not treated as strict ground truth because a small fraction of originally missing values are interpolated.

Samples are retained when valid pixels with reflectivity exceeding 20 dBZ at 1.0 km occupy at least 10\% of the 256 $\times$ 256 domain, yielding 21,919 samples. These are divided chronologically into 17,535 training samples from January 2023 to 6 June 2025 and 4,384 test samples from 7 June to 31 December 2025. Structured synthetic missingness is applied to the training samples as described in Section III-B. RadarFill and 3D U-Net-GAN use the same CAP training and test partitions.

\begingroup
\footnotesize
\setlength{\tabcolsep}{3pt}
\begin{longtable}{@{}>{\raggedright\arraybackslash}p{0.2701\linewidth}>{\raggedright\arraybackslash}p{0.2701\linewidth}>{\raggedright\arraybackslash}p{0.2004\linewidth}>{\raggedright\arraybackslash}p{0.1995\linewidth}@{}}
\caption{CAP RADAR-DATA COMPLETENESS STATISTICS IN THE TRAINING REGION (118.00$^\circ$E--120.56$^\circ$E, 31.00$^\circ$N--33.56$^\circ$N)}\\

\toprule\noalign{}
\begin{minipage}[b]{\linewidth}\raggedright
\textbf{Level}
\end{minipage} & \begin{minipage}[b]{\linewidth}\raggedright
\textbf{Altitude (km)}
\end{minipage} & \begin{minipage}[b]{\linewidth}\raggedright
\textbf{Mean missing fraction}
\end{minipage} & \begin{minipage}[b]{\linewidth}\raggedright
\textbf{Valid sample count}
\end{minipage} \\
\midrule\noalign{}
\endhead
\bottomrule\noalign{}
\endlastfoot
1 & 1.0 & 2.0928\% & 17534 \\
2 & 1.5 & 0.1501\% & 17527 \\
3 & 2.0 & 0.0200\% & 17532 \\
4 & 2.5 & 0.0005\% & 17529 \\
5 & 3.0 & 0.0007\% & 17530 \\
6 & 3.5 & 0.0006\% & 17530 \\
7 & 4.0 & 0.0004\% & 17532 \\
8 & 4.5 & 0.0004\% & 17528 \\
9 & 5.0 & 0.0005\% & 17533 \\
10 & 5.5 & 0.0005\% & 17528 \\
11 & 6.0 & 0.0005\% & 17529 \\
12 & 6.5 & 0.0005\% & 17529 \\
13 & 7.0 & 0.0005\% & 17532 \\
14 & 7.5 & 0.0005\% & 17530 \\
15 & 8.0 & 0.0006\% & 17533 \\
16 & 9.0 & 0.0006\% & 17527 \\
17 & 10.0 & 0.0007\% & 17533 \\
18 & 11.0 & 0.0007\% & 17529 \\
19 & 12.0 & 0.0008\% & 17530 \\
20 & 13.0 & 0.0009\% & 17528 \\
21 & 14.0 & 0.0010\% & 17533 \\
22 & 15.0 & 0.0014\% & 17529 \\
23 & 16.0 & 0.0020\% & 17531 \\
23-level mean & --- & 0.0990\% & --- \\
\end{longtable}
\endgroup

Note: Missing fractions are calculated from the original validity masks for each time and altitude; the table reports the training-set mean at each altitude. The 23-level mean is the arithmetic mean of the layer-wise missing fractions across all modeled altitudes. The low overall fraction of original missing data indicates that linear pre-interpolation affects only a small fraction of voxels in the observation field, with the largest influence at the lowest levels.

\FloatBarrier

\subsection{Minhou Phased-Array Radar Data}

Independent validation used observations from an S-band dual-polarization phased-array weather radar in Minhou County, Fujian Province (26.3910$^\circ$N, 119.0340$^\circ$E), during Typhoon Doksuri on 28 July 2023. The radar scans 23 elevation angles from 0$^\circ$ to 20$^\circ$, with a dual-polarization observation range of approximately 230 km, a radial resolution of 31.25 m, and minute-scale volume scans. Because the Minhou radar does not contribute to the CAP V3.0 mosaic product, its observations provide an independent reference for evaluating reconstructed reflectivity within real CAP blind zones.

The phased-array volumes were quality controlled and converted to constant-altitude plan position indicator (CAPPI) reflectivity fields. At each target altitude, valid observations within $\pm$0.5 km were interpolated in two dimensions to generate CAPPI fields from 1 to 16 km and then resampled to the 0.01$^\circ$ $\times$ 0.01$^\circ$ grid used by CAP. No extrapolation was applied beyond the valid phased-array observation area. CAP and phased-array fields were matched by observation time with a maximum offset of 1 min, yielding 90 matched volume scans over the validation domain (24.44$^\circ$--27.00$^\circ$N, 117.44$^\circ$--120.00$^\circ$E).

Because CAP is a multi-radar mosaic whereas the Minhou dataset is obtained from a single-site radar, the two products are not expected to agree exactly. Their observational consistency was therefore characterized at common valid grid points using CAP-minus-phased-array bias and root-mean-square error (RMSE) (Table II). This comparison is descriptive only and is not used for model training or reflectivity correction.

\begingroup
\footnotesize
\setlength{\tabcolsep}{3pt}
\begin{longtable}{@{}>{\raggedright\arraybackslash}p{0.2701\linewidth}>{\raggedright\arraybackslash}p{0.2701\linewidth}>{\raggedright\arraybackslash}p{0.2004\linewidth}>{\raggedright\arraybackslash}p{0.1995\linewidth}@{}}
\caption{LAYER-WISE REFLECTIVITY CONSISTENCY BETWEEN CAP AND THE MINHOU PHASED-ARRAY RADAR}\\

\toprule\noalign{}
\begin{minipage}[b]{\linewidth}\raggedright
\textbf{Altitude (km)}
\end{minipage} & \begin{minipage}[b]{\linewidth}\raggedright
\textbf{Common valid matched grid points}
\end{minipage} & \begin{minipage}[b]{\linewidth}\raggedright
\textbf{Bias (dBZ)}
\end{minipage} & \begin{minipage}[b]{\linewidth}\raggedright
\textbf{RMSE (dBZ)}
\end{minipage} \\
\midrule\noalign{}
\endhead
\bottomrule\noalign{}
\endlastfoot
1.0 & 1,550,503 & $-$2.97 & 13.77 \\
1.5 & 2,880,029 & 0.21 & 12.81 \\
2.0 & 3,917,846 & $-$0.11 & 12.24 \\
2.5 & 4,651,097 & 1.09 & 10.56 \\
\end{longtable}
\endgroup

\FloatBarrier

\section{RadarFill Method}

\FloatBarrier

\subsection{Observation-Constrained Conditional Diffusion Framework}

RadarFill formulates gap filling in systematic 3-D radar observation blind zones as a conditional generation problem constrained by available valid radar observations. Let X denote the complete 23-level reflectivity field and \(\text{X}_{\text{obs}}\) the reflectivity field after missingness is imposed. Let M be the binary observation mask, where M = 1 denotes a valid observation and M = 0 denotes a location to be filled. The model conditions on \(\text{X}_{\text{obs}}\) and M and obtains a generated reflectivity field \(\text{X}_{\text{gen}}\) through diffusion sampling. Only generated values at missing locations are retained and combined with the original valid observations to form the complete reflectivity field:

\begin{equation}
X_{\mathrm{comp}}=M\odot X_{\mathrm{obs}}+(1-M)\odot X_{\mathrm{gen}}
\label{eq:3}
\end{equation}
where \(\odot\) denotes element-wise multiplication. Thus, original measurements are preserved wherever M=1, whereas model-generated values are used only within the missing region. The overall observation-constrained conditional diffusion framework is illustrated in Fig. 1, and the implementation of the observational constraints during reverse sampling is described in Section III-D.

\begin{figure}[H]
\centering
\includegraphics[width=\linewidth,height=0.73\textheight,keepaspectratio]{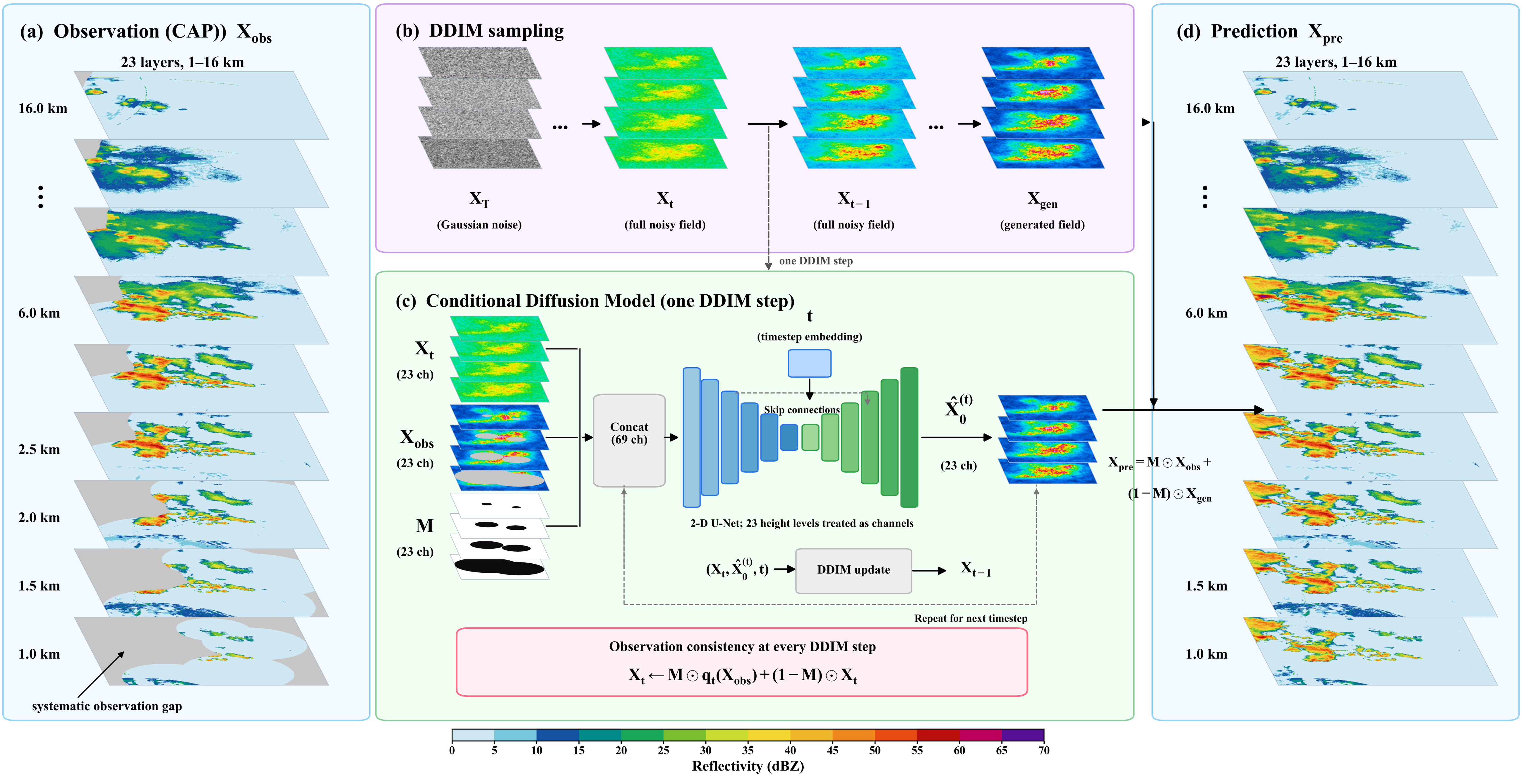}
\caption{Observation-constrained conditional diffusion framework of RadarFill.}\label{fig:1}
\end{figure}

\FloatBarrier

\subsection{Construction of Structured Synthetic Missingness}

To support model training and controlled evaluation, structured missing regions are synthesized from relatively complete CAP observations. Layer-wise missing fractions were first calculated over Shanghai and Zhejiang. Although the absolute missing fractions differ between the two regions, both show a common vertical pattern of larger missingness at low levels and a rapid decrease with altitude (Table III). Based on this pattern, the target missing fractions at 1.0, 1.5, 2.0, and 2.5 km are set to 50\%, 30\%, 20\%, and 10\%, respectively, and generally decrease with altitude. These target fractions are intended to represent relatively severe low-level missingness rather than reproduce the missing-data distribution of a specific region.

We further quantified the spatial containment of missing regions between adjacent altitude levels over Zhejiang and the area surrounding Shanghai. Within 1.0--2.5 km, approximately 77.69\%--99.97\% of missing pixels at a higher level also fall within the missing region at the adjacent lower level, indicating pronounced vertical nesting in the CAP data. Complete layer-wise statistics are provided in Supplementary Table S1.

For each sample, 2--5 spatial centers are randomly generated, with a reference scale drawn from 12\% to 30\% of the horizontal domain size. A common spatial ordering is then defined from the normalized distance to these centers, and the prescribed fraction of pixels at each altitude is selected as missing according to this ordering. This produces missing regions that are broader at low levels and progressively contract with altitude. Their locations and scales are randomized during training to provide a range of structured missingness patterns. This construction is used as a data-informed experimental design that preserves the principal vertical organization of systematic low-level blind zones rather than as a statistical model of regional missingness.

\begingroup
\footnotesize
\setlength{\tabcolsep}{3pt}
\begin{longtable}{@{}>{\raggedright\arraybackslash}p{0.2785\linewidth}>{\raggedright\arraybackslash}p{0.1529\linewidth}>{\raggedright\arraybackslash}p{0.2159\linewidth}>{\raggedright\arraybackslash}p{0.2927\linewidth}@{}}
\caption{LAYER-WISE CAP MISSING FRACTIONS OVER SHANGHAI AND ZHEJIANG AND TARGET FRACTIONS FOR STRUCTURED SYNTHETIC MISSINGNESS}\\

\toprule\noalign{}
\begin{minipage}[b]{\linewidth}\raggedright
\textbf{Altitude}
\end{minipage} & \begin{minipage}[b]{\linewidth}\raggedright
\textbf{Shanghai missing fraction (\%)}
\end{minipage} & \begin{minipage}[b]{\linewidth}\raggedright
\textbf{Zhejiang missing fraction (\%)}
\end{minipage} & \begin{minipage}[b]{\linewidth}\raggedright
\textbf{Target missing fraction for structured synthetic missingness (\%)}
\end{minipage} \\
\midrule\noalign{}
\endhead
\bottomrule\noalign{}
\endlastfoot
1.0 km & 22.12 & 48.19 & 50.0 \\
1.5 km & 0.64 & 17.99 & 30.0 \\
2.0 km & 0.45 & 5.57 & 20.0 \\
2.5 km & 0.09 & 2.49 & 10.0 \\
3.0 km & 0.09 & 1.00 & 10.0 \\
3.5 km & 0.00 & 0.24 & 9.5 \\
4.0 km & 0.00 & 0.15 & 9.0 \\
4.5 km & 0.00 & 0.10 & 8.5 \\
5.0 km & 0.00 & 0.05 & 8.0 \\
5.5 km & 0.00 & 0.03 & 7.5 \\
6.0 km & 0.00 & 0.03 & 7.0 \\
6.5 km & 0.00 & 0.01 & 6.5 \\
7.0 km & 0.00 & 0.01 & 6.0 \\
7.5 km & 0.00 & 0.01 & 5.5 \\
8.0 km & 0.00 & 0.00 & 5.0 \\
9.0 km & 0.00 & 0.00 & 4.5 \\
10.0 km & 0.00 & 0.00 & 4.0 \\
11.0 km & 0.00 & 0.00 & 3.5 \\
12.0 km & 0.00 & 0.00 & 3.0 \\
13.0 km & 0.00 & 0.00 & 2.5 \\
14.0 km & 0.02 & 0.00 & 2.0 \\
15.0 km & 0.67 & 0.01 & 1.5 \\
16.0 km & 1.08 & 0.10 & 1.0 \\
\end{longtable}
\endgroup

Note: Actual missing fractions in CAP mosaic products vary substantially by region. The Shanghai and Zhejiang statistics are used only to characterize the common vertical pattern of greater low-level missingness that generally decreases with altitude, rather than to construct an empirical model of regional missing fractions. Target fractions for structured synthetic missingness are deliberately set to represent relatively severe missingness, increasing task difficulty and covering possible scenarios with more extensive low-level observation gaps. They should therefore not be interpreted as fits to the actual missing fraction in Shanghai, Zhejiang or any other specific region.

\FloatBarrier

\subsection{Conditional Diffusion Model}

RadarFill performs diffusion directly in the normalized reflectivity space without latent encoding. The forward process follows a denoising diffusion probabilistic model (DDPM) \cite{ref26} with T=1000 time steps. The variance schedule is defined as

\begin{equation}
\beta_t=\left[\sqrt{\beta_{\mathrm{start}}}+\frac{t}{T-1}\left(\sqrt{\beta_{\mathrm{end}}}-\sqrt{\beta_{\mathrm{start}}}\right)\right]^2
\label{eq:4}
\end{equation}
Where \(\beta_{\text{start}}\text{=0.0015,}\beta_{\text{end}}\text{=0.0155}\), and \(\text{t}\text{=0,\ldots{},}\text{T}\text{$-$1}\). Defining

\begin{equation}
\alpha_t=1-\beta_t
\label{eq:5}
\end{equation}
\begin{equation}
\bar\alpha_t=\prod_{s=0}^{t}\alpha_s
\label{eq:6}
\end{equation}
the noisy reflectivity field at time step t is obtained as

\begin{equation}
X_t=\sqrt{\bar\alpha_t}\,X_0+\sqrt{1-\bar\alpha_t}\,\epsilon,\qquad \epsilon\sim\mathcal N(0,I)
\label{eq:7}
\end{equation}
During training, (t) is uniformly sampled from \(\text{0,\ldots{},999}\). RadarFill uses \(\text{x}_{\text{0}}\)-parameterization: the network directly predicts the corresponding noise-free reflectivity field from the current diffusion state and the conditioning information rather than predicting the added noise.

The 23 altitude levels from 1 to 16 km are treated as channels of a 2-D U-Net rather than processed with explicit 3-D convolutions. The conditioning information comprises the 23-channel incomplete normalized reflectivity field and the corresponding 23-channel observation mask. Missing locations in the normalized reflectivity field are filled with (-1.0) and explicitly identified by M=0, distinguishing them from valid low-reflectivity observations. At each diffusion step, the 23-channel diffusion state, 23-channel incomplete reflectivity field, and 23-channel mask are concatenated to form a 69-channel input, and the network outputs a 23-channel estimate of the noise-free reflectivity field.

The U-Net uses 128 base channels with channel multipliers [1,2,2,4,4] and two residual blocks at each resolution level. Spatial attention is applied at the \(\text{32$\times$32}\) and \(\text{16$\times$16}\) feature scales, using 32 channels per attention head. The diffusion time step is embedded into 128 dimensions and projected to 512 dimensions through a two-layer multilayer perceptron (MLP) with SiLU activation before being injected into the residual blocks.

Training uses a spatially weighted \(\text{L}_{\text{2}}\) loss to emphasize reconstruction within the synthetically missing region. For voxel i,

\begin{equation}
\mathcal L=\frac{\sum_i w_i(\hat X_{0,i}-X_{0,i})^2}{\sum_i w_i},\qquad w_i=\begin{cases}1.0,&M_i=0,\\0.02,&M_i=1.\end{cases}
\label{eq:8}
\end{equation}
where missing and observed voxels are assigned weights of 1.0 and 0.02, respectively.

\FloatBarrier

\subsection{Model Training and Inference}

RadarFill is implemented in PyTorch Lightning and optimized using AdamW \cite{ref27} with an initial learning rate of \(\text{5$\times$1}\text{0}^{\text{$-$5}}\), a batch size of 8 and 120 training epochs. Gradient clipping is applied with a threshold of 1.0, and the training random seed is fixed at 23. An exponential moving average (EMA) of the model parameters is updated after each training batch, and the EMA weights from the final 120-epoch checkpoint are used for inference. Experiments were conducted on NVIDIA A100-PCIE-40GB GPUs using Python 3.8.20, PyTorch 2.0.1 with CUDA 11.7, Diffusers 0.25.1, and Accelerate 0.27.2.

Inference uses 200-step denoising diffusion implicit model (DDIM) sampling \cite{ref28} with \(\eta\text{=1.0}\). The sampling time steps are uniformly subsampled from the 1,000-step diffusion process, corresponding to indices (\(\text{1,6,\ldots{},996}\)), and traversed in reverse order during sampling. A single realization is generated for each input without ensemble averaging, using a fixed random seed of 42. Sampling is initialized from Gaussian noise, while the incomplete reflectivity field and observation mask remain fixed conditioning inputs throughout the reverse process.

To maintain consistency with available observations, at each reverse time step (t), the observed reflectivity is first mapped to the corresponding noise level,

\begin{equation}
q_t(X_{\mathrm{obs}})=\sqrt{\bar\alpha_t}\,X_{\mathrm{obs}}+\sqrt{1-\bar\alpha_t}\,\epsilon_t,\qquad\epsilon_t\sim\mathcal N(0,I)
\label{eq:9}
\end{equation}
where \(\epsilon_{\text{t}}\) is resampled at each reverse time step. The known locations in the current diffusion state are then replaced according to

\begin{equation}
\tilde X_t=M\odot q_t(X_{\mathrm{obs}})+(1-M)\odot X_t
\label{eq:10}
\end{equation}
Thus, observed locations remain constrained by the original measurements at the corresponding noise level throughout reverse sampling, whereas the current generative state is retained at missing locations. After sampling, the generated field is combined with the original observations according to \eqref{eq:3}, so that valid observations are preserved exactly and generated values are used only where M=0. The completed reflectivity field is finally transformed back to physical units and clipped to 0--70 dBZ. Inference takes approximately 15 s per 256 $\times$ 256 $\times$ 23 radar volume using 200-step DDIM sampling on an NVIDIA A100 GPU.

\FloatBarrier

\section{Experimental Design and Statistical Evaluation}

\FloatBarrier

\subsection{Synthetic-Missingness Evaluation}

Synthetic-missingness evaluation was conducted over the 4,384 held-out test times and representative thunderstorm and typhoon cases. RadarFill was compared with fill\_data\_3D, the gap-filling procedure used in the Nowcast3D preprocessing workflow, and the learning-based 3D U-Net-GAN. fill\_data\_3D fills low-level missing voxels supported by upper-level echoes using ridge regression on higher-level reflectivity followed by guided-filter refinement, while preserving valid observations. All methods used the same test samples, synthetic-missingness masks, evaluation regions, and metrics.

The 3D U-Net-GAN reference was independently implemented following the architecture described for Rad3DGFM \cite{ref23}, using a 3D U-Net with inter-layer attention and a conditional least-squares GAN. The input and output were adapted to the 23 altitude levels from 1 to 16 km, with corresponding adjustments to the vertical downsampling and upsampling scales in the encoder and decoder. The model was trained on the same 17,535 CAP samples used for RadarFill. Its generator loss combines mean absolute error (MAE), mean squared error (MSE), and adversarial terms with weights of 0.6, 0.3, and 0.1, respectively. Training used a physical batch size of 2 with gradient accumulation over eight steps, with generator and discriminator learning rates of ~$2\times10^{-4}$ and $5\times10^{-4}$, respectively. The test set was not used for model tuning or checkpoint selection.

Gap-filling performance was evaluated using RMSE, bias, critical success index (CSI), and fractions skill score (FSS) \cite{ref29}, together with composite reflectivity, echo-top height, and the 3-D reflectivity-weighted centroid. For the representative thunderstorm case, echo-top height and the 3-D reflectivity-weighted centroid were evaluated at 35, 45, and 55 dBZ within the synthetically masked region. Echo-top height was defined as the highest altitude exceeding the threshold in each horizontal column, and the centroid was calculated over threshold-exceeding voxels using reflectivity in dBZ as the weight. RMSE and bias were calculated only within the synthetically masked region for voxels whose observation-field reflectivity reached the corresponding threshold. For reflectivity threshold T, RMSE and bias are defined as

\begin{equation}
\mathrm{RMSE}_T=\sqrt{\frac{1}{N_T}\sum_{i:y_i\geq T}(\hat y_i-y_i)^2}
\label{eq:11}
\end{equation}
\begin{equation}
\mathrm{Bias}_T=\frac{1}{N_T}\sum_{i:y_i\geq T}(\hat y_i-y_i)
\label{eq:12}
\end{equation}
where \(\text{y}_{\text{i}}\) and \({\widehat{\text{y}}}_{\text{i}}\) denote reflectivity in the observation and gap-filled fields, respectively, and \(\text{N}_{\text{T}}\) is the number of valid voxels within the synthetically masked region for which the observation-field reflectivity is not lower than threshold T.

CSI was used to evaluate echo detection at different reflectivity thresholds and was calculated over the entire synthetically masked region without applying the observation-threshold screening used for RMSE and bias:

\begin{equation}
\mathrm{CSI}=\frac{\mathrm{TP}}{\mathrm{TP}+\mathrm{FP}+\mathrm{FN}}
\label{eq:13}
\end{equation}
where TP, FP and FN denote the numbers of hits, false alarms and misses, respectively.

FSS was used to assess spatial consistency at thresholds of 15, 25, and 35 dBZ over neighborhood scales of approximately 1, 5, 10, 20, 40, and 80 km, with statistics jointly calculated over the 1--8-km altitude range:

\begin{equation}
\mathrm{FSS}=1-\frac{\sum (f_p-f_o)^2}{\sum (f_p^2+f_o^2)}
\label{eq:14}
\end{equation}
where \(\text{f}_{\text{p}}\) and \(\text{f}_{\text{0}}\) denote the fractions of pixels reaching the corresponding reflectivity threshold within each neighborhood in the gap-filled and observation fields, respectively. FSS was restricted to the synthetically masked region, with neighborhood centers located within the mask, and only neighborhoods with at least 50\% valid pixels within the masked region were retained. For the large-sample evaluation, shading and error bars in Fig. 4 represent the 25th--75th percentile range of the corresponding sample-wise metric distributions.

\FloatBarrier

\subsection{Independent Validation in Real Observation Blind Zones}

Independent validation was performed using the 90 matched CAP--phased-array volume scans described in Section II-B. Evaluation was restricted to the intersection of the original CAP blind zones, valid phased-array observations, and locations where both gap-filled products were available.

Validation focused on low-level reflectivity recovery, reflectivity distributions, and composite reflectivity. The primary quantitative analysis pooled missing voxels at 1.0, 1.5, and 2.0 km, which account for approximately 99.25\% of the common valid missing voxels over 1.0--2.5 km. RMSE and CSI were calculated using the definitions in Section IV-A, with phased-array reflectivity as the observational reference: RMSE was evaluated only where phased-array reflectivity reached the corresponding threshold, whereas CSI was calculated over all common valid missing voxels. Reflectivity distributions were evaluated over common valid samples from 1.0--2.5 km with phased-array reflectivity \(\text{$\geq$15}\) dBZ. Composite reflectivity was defined as the maximum reflectivity over vertical levels valid in both the gap-filled field and phased-array observations and was evaluated over the horizontal extent of the low-level blind zone.

To give each matched volume scan equal weight, RMSE and CSI were first calculated separately for each scan and then averaged across valid times. Because consecutive scans are temporally correlated, 95\% confidence intervals were estimated using moving-block bootstrap resampling. Based on the temporal autocorrelation of the metric series, a block length of 18 scans (approximately 108 min) was used with 10,000 resamples. Identical resampled time blocks were applied to the two gap-filling methods.

\FloatBarrier

\subsection{Nowcasting Experiment}

The nowcasting experiment used the open-source pretrained Nowcast3D model to test how 3-D reflectivity initial fields produced by different gap-filling methods affect subsequent forecasts. We selected a typhoon event on 30 July 2025 and a precipitation event on 25 August 2025, each comprising 201 forecast cycles. For both events, structured synthetic missingness was imposed on relatively complete CAP observation fields. Following the Nowcast3D input configuration, each initialization used 10 consecutive volume scans as the historical input sequence. The same synthetic missingness mask was applied to all 10 historical scans within each forecast initialization. Because adjacent forecast cycles within the same event share part of their historical input, the 201 cycles cannot be treated as independent weather samples. The two events were evaluated separately, without pooling their statistics.

Three experiments, denoted Fill, RadarFill, and Observation, use the fill\_data\_3D gap-filled fields, RadarFill gap-filled fields, and observation fields without synthetic missingness, respectively, as historical inputs. The downstream comparison focuses on fill\_data\_3D because it is the gap-filling procedure used in the original Nowcast3D preprocessing workflow, allowing the impact of replacing the existing gap-filling step with RadarFill to be assessed directly. All experiments use the same pretrained Nowcast3D model without retraining or parameter adjustment and identical forecast settings, with lead times from 6 to 180 min. Forecasts are evaluated against the corresponding observation fields, with Observation serving as a reference using unmasked historical inputs. Because Nowcast3D requires a 0.5-km level whereas RadarFill operates over 1--16 km, the 1.0-km field is copied to 0.5 km for all three experiments to form the required 24-level input. Quantitative evaluation begins at 1.0 km and excludes the copied 0.5-km level.

Forecast performance is evaluated using RMSE, bias, and CSI at reflectivity thresholds of 15, 25, and 35 dBZ, together with altitude--lead-time variations in RMSE. Results for the typhoon event are presented in the main text, and those for the 25 August precipitation event are provided in the Supplementary Material. The relative CSI improvement of RadarFill over Fill is defined as

\begin{equation}
I_{\mathrm{CSI}}=\frac{\mathrm{CSI}_{\mathrm{RadarFill}}-\mathrm{CSI}_{\mathrm{Fill}}}{\mathrm{CSI}_{\mathrm{Fill}}}\times 100\%
\label{eq:15}
\end{equation}
The corresponding relative RMSE improvement is

\begin{equation}
I_{\mathrm{RMSE}}=\frac{\mathrm{RMSE}_{\mathrm{Fill}}-\mathrm{RMSE}_{\mathrm{RadarFill}}}{\mathrm{RMSE}_{\mathrm{Fill}}}\times 100\%
\label{eq:16}
\end{equation}
and the relative improvement in bias magnitude is

\begin{equation}
I_{\mathrm{Bias}}=\frac{|\mathrm{Bias}_{\mathrm{Fill}}|-|\mathrm{Bias}_{\mathrm{RadarFill}}|}{|\mathrm{Bias}_{\mathrm{Fill}}|}\times 100\%
\label{eq:17}
\end{equation}
Because the forecast model and settings are held fixed, these experiments isolate the influence of the different 3-D reflectivity initial fields on subsequent nowcasts.

\FloatBarrier

\section{Results}

\FloatBarrier

\subsection{Recovery of High-Reflectivity Convective Structures Under Extensive Low-Level Missingness}

We first examined the recovery of high-reflectivity convective structures under extensive low-level missingness using the thunderstorm case on 25 August 2025. The missing fractions at 1.0, 1.5, 2.0, and 2.5 km were 50\%, 30\%, 20\%, and 10\%, respectively. Within the 1.0--2.5-km observation field, 66.8\%, 69.6\%, and 53.1\% of echoes\(\text{ }\text{$\geq$}\)35, \(\text{$\geq$}\)45, and \(\text{$\geq$55}\) dBZ, respectively, were located within the synthetically masked region. Despite the removal of a large fraction of the low-level strong echoes, RadarFill recovered the principal high-reflectivity structures within the missing region. The upper tails of the reflectivity distributions in the observation and gap-filled fields were similar (Fig. 2a). Layer-wise CSI was generally high at 15 and 25 dBZ and remained relatively high at most altitudes at 35 dBZ, whereas lower values and greater interlevel variability occurred at 45 and 55 dBZ, particularly at 55 dBZ (Fig. 2b).

The recovered fields were further examined in terms of the vertical profile of reflectivity (VPR) and spatial structure. The VPR around the strong-echo core broadly reproduced the vertical intensity distribution of the deep echo column in the observation field, with a modest positive bias at approximately 4--8 km (Fig. 2c). On the 1.0--2.5-km constant-altitude planes and in composite reflectivity, the locations and morphology of the principal convective cells within the synthetically masked region were well recovered. The zonal and meridional cross-sections likewise recovered the principal low-level strong echoes, with their horizontal positions and vertical extents broadly consistent with those in the observation field. The larger local deviations were concentrated near the center of the synthetically masked region (Fig. 2f).

Echo-top height and the 3-D reflectivity-weighted centroid were used to quantify differences in spatial structure (Fig. 2d,e). At thresholds of 35, 45, and 55 dBZ, the horizontal centroid displacements were 1.96, 3.62, and 6.96 km, respectively, and the corresponding vertical displacements were $-$0.23, $-$0.08, and $-$0.22 km. The horizontal displacement was largest at 55 dBZ, whereas the vertical displacement remained below 0.25 km at all three thresholds. The corresponding mean echo-top-height biases were $-$0.28, $-$0.14, and $-$0.12 km, all smaller in magnitude than the 0.5-km vertical spacing of CAP over 1--8 km; the differences in maximum echo-top height were $-$1, 0, and $-$1 km, respectively. These results show that RadarFill recovered the principal high-reflectivity regions and their 3-D structures within the extensive low-level missing region in this case.

\begin{figure}[H]
\centering
\includegraphics[width=\linewidth,height=0.73\textheight,keepaspectratio]{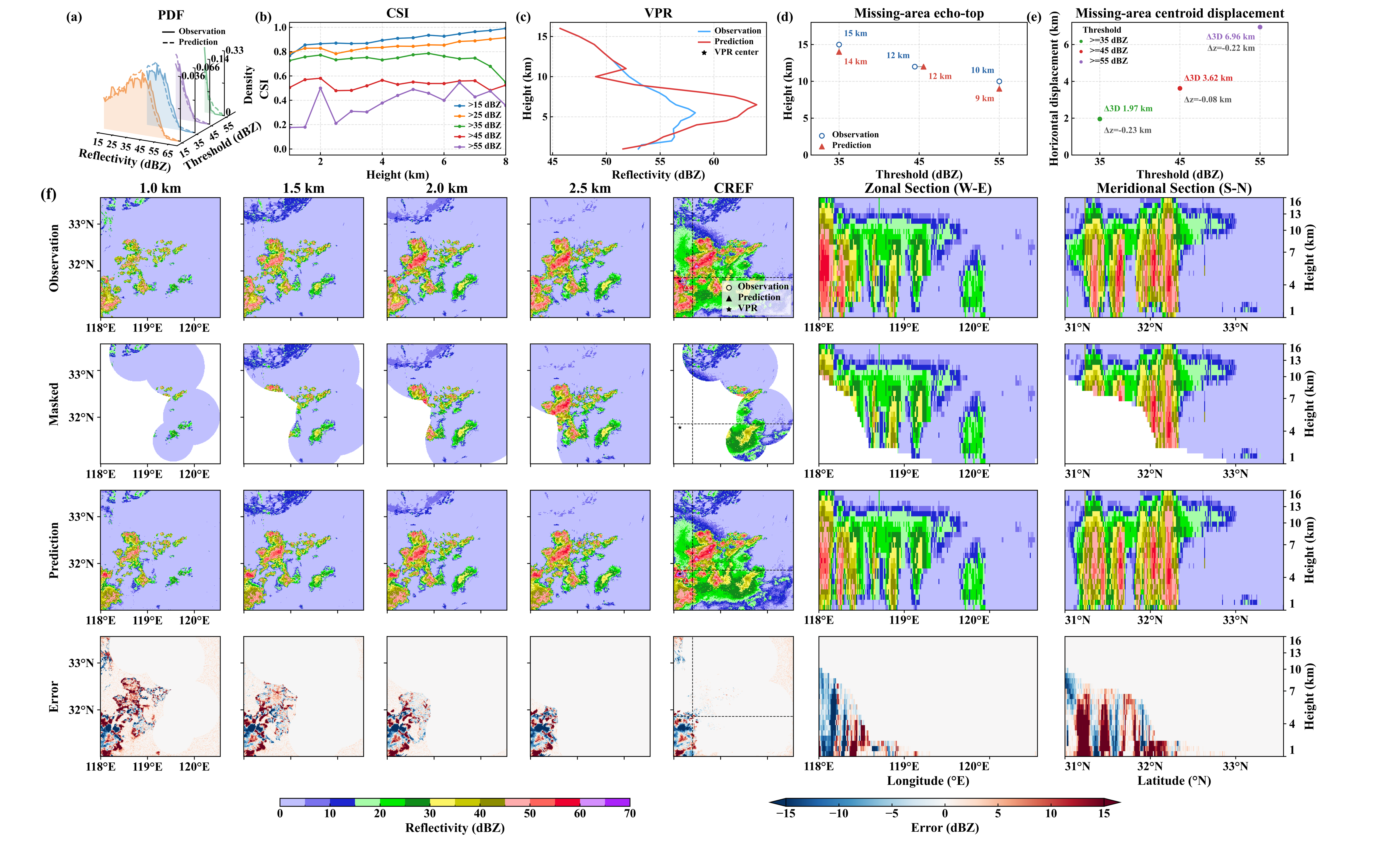}
\caption{Gap filling of the 3-D structure of strong thunderstorm echoes under extreme low-level missingness. (a) Probability density functions (PDFs) of the observation and RadarFill gap-filled fields within the synthetically masked region at different reflectivity thresholds. (b) Layer-wise critical success index (CSI) within the synthetically masked region at thresholds of 15, 25, 35, 45, and 55 dBZ. (c) Vertical profile of reflectivity (VPR) around the strong-echo core. The center is defined as the location with the maximum 4 $\times$ 4 neighborhood-mean reflectivity within the low-level synthetically masked region, and the VPR is the mean reflectivity over a 5 $\times$ 5 grid surrounding this location. (d) Maximum echo-top height within the synthetically masked region at thresholds of 35, 45, and 55 dBZ. (e) Displacement of the 3-D reflectivity-weighted centroid within the synthetically masked region at thresholds of 35, 45, and 55 dBZ. The vertical axis represents horizontal displacement; vertical displacement ($\Delta$z) and total 3D displacement ($\Delta$3D) are also annotated. (f) Constant-altitude planes at 1.0, 1.5, 2.0, and 2.5 km, composite reflectivity (CREF), and zonal (W--E) and meridional (S--N) vertical cross-sections along 31.86$^\circ$N and 118.40$^\circ$E, respectively. From top to bottom are the observation field, missing-data input, RadarFill gap-filled field, and error field. Circles and triangles in the composite-reflectivity panels denote the reflectivity-weighted centroids of the observation and gap-filled fields, respectively, and the star denotes the VPR center. Dashed lines indicate the zonal and meridional cross-sections, which intersect at 118.40$^\circ$E, 31.86$^\circ$N. Except for the VPR, the statistics in (a), (b), (d), and (e) are restricted to the synthetically masked region. Error is defined as Prediction $-$ Observation; positive values indicate higher reflectivity in the gap-filled field than in the observation field, and negative values indicate lower reflectivity.}\label{fig:2}
\end{figure}

\FloatBarrier

\subsection{Gap Filling of Large-Scale Typhoon Rainband Structures}

We further examined RadarFill for large-scale typhoon precipitation structures using Typhoon Co-May on 31 July 2025. To cover the main precipitation region, a 512 $\times$ 512 domain spanning 116.00$^\circ$E--121.12$^\circ$E and 29.00$^\circ$N--34.12$^\circ$N was divided into four 256 $\times$ 256 subdomains for separate inference. The typhoon exhibited well-defined spiral rainbands, while synthetic missingness removed large portions of several low-level rainbands. After gap filling, the principal spiral rainbands were recovered on the 1.0--2.5-km constant-altitude planes and in composite reflectivity, with locations, morphology, and spatial continuity broadly consistent with the observation field (Fig. 3d).

Within the synthetically masked region over 1--8 km, FSS at 15 and 25 dBZ increased from 0.695 and 0.615, respectively, at the 1-km neighborhood scale to 0.972 and 0.961 at 80 km (Fig. 3a). At 35 dBZ, FSS was lower at small neighborhood scales but increased markedly with scale, indicating greater difficulty in the local positioning of stronger echoes while their larger-scale spatial distribution remained more consistent with the observation field. Gap-filled reflectivity was slightly lower than the observation field overall, with layer-wise mean bias ranging from approximately $-$2.0 to $-$5.3 dBZ over 1--8 km (Fig. 3b). Layer-wise CSI also varied with threshold and altitude, ranging approximately from 0.40--0.68, 0.40--0.62, and 0.29--0.67 at 15, 25, and 35 dBZ, respectively, at levels with valid statistics (Fig. 3c). At 35 dBZ, strong-echo samples within the masked region were sparse above 6 km, leaving CSI undefined at some levels; these upper-level values are therefore not further interpreted.

The zonal (W--E) and meridional (S--N) cross-sections through the outer and inner spiral rainbands further show that RadarFill recovered the principal echo structures of both rainbands within the synthetically masked region (Fig. 3d). Their horizontal positions and vertical extents were broadly consistent with the observation field, while the larger local deviations were concentrated near the center of the masked region. Overall, RadarFill recovered the principal 3-D spatial distribution and vertical structure of the typhoon spiral rainbands in this case.

\begin{figure}[H]
\centering
\includegraphics[width=\linewidth,height=0.73\textheight,keepaspectratio]{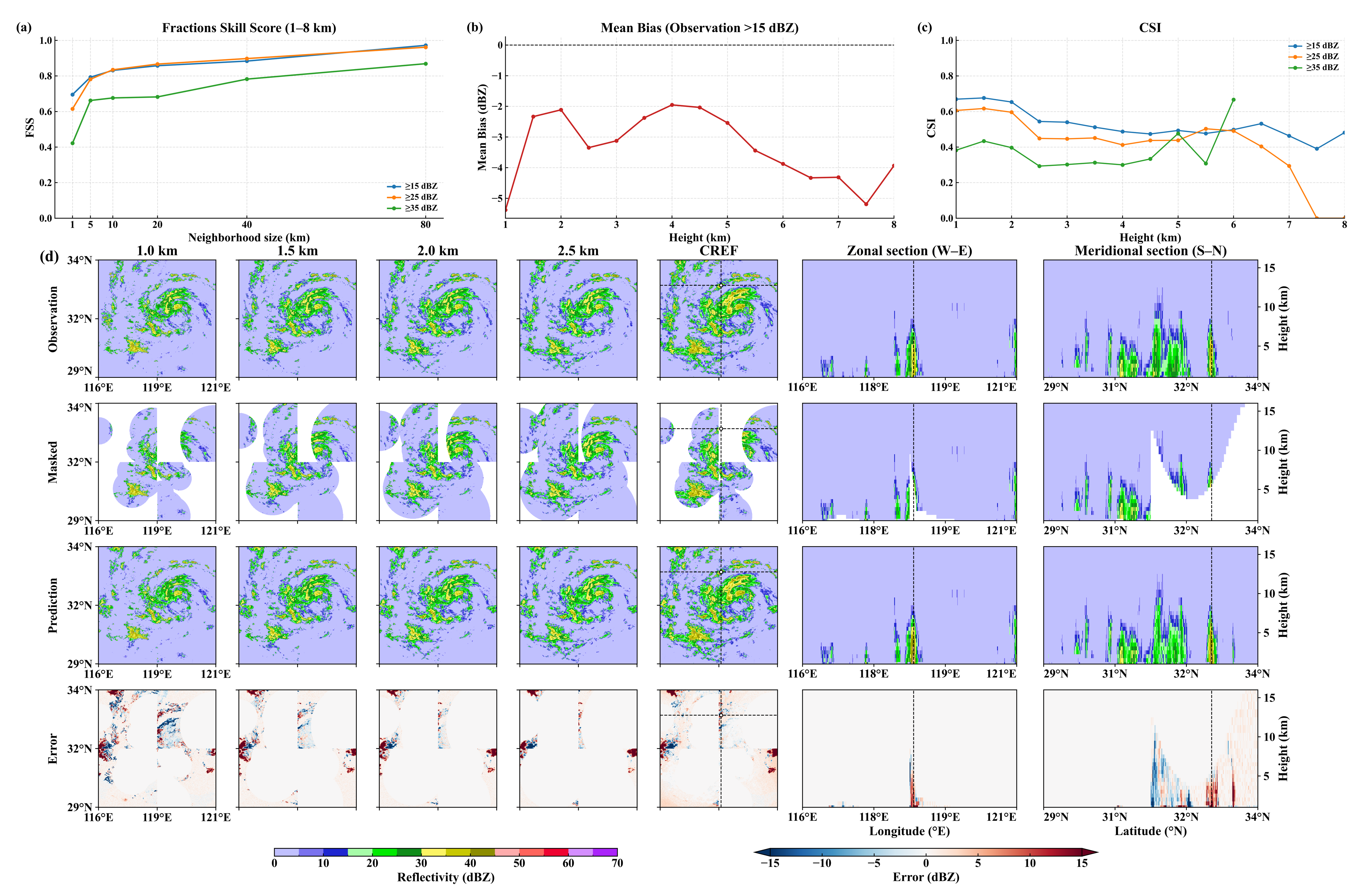}
\caption{Gap filling of typhoon precipitation structures under extensive low-level missingness. The domain spans 116.00$^\circ$E--121.12$^\circ$E and 29.00$^\circ$N--34.12$^\circ$N on a 512 $\times$ 512 horizontal grid, obtained by independently inferring four 256 $\times$ 256 subdomains and mosaicking the results. (a) Fractions Skill Score (FSS) as a function of neighborhood scale, jointly evaluated over 1--8 km within the synthetically masked region at thresholds of 15, 25 and 35 dBZ. (b) Layer-wise mean Bias for samples with observed reflectivity $\geq$15 dBZ within the synthetically masked region. (c) Layer-wise critical success index (CSI) within the synthetically masked region at thresholds of 15, 25 and 35 dBZ. (d) Constant-altitude planes at 1.0, 1.5, 2.0 and 2.5 km, composite reflectivity (CREF), and zonal (W--E) and meridional (S--N) vertical cross-sections along 33.02$^\circ$N and 118.66$^\circ$E, respectively. The zonal section crosses the outer spiral rainband and the meridional section crosses the inner spiral rainband. From top to bottom: observation field, missing-data input, RadarFill gap-filled field and error field. Dashed lines in the composite-reflectivity panels mark the two vertical cross-sections, which intersect at 118.66$^\circ$E, 33.02$^\circ$N. Bias is calculated only for samples with observed reflectivity $\geq$15 dBZ within the synthetically masked region; CSI is calculated over the full synthetically masked region at the corresponding reflectivity threshold. Error is defined as Prediction $-$ Observation; positive values indicate higher reflectivity in the gap-filled field and negative values indicate lower reflectivity.}\label{fig:3}
\end{figure}

\FloatBarrier

\subsection{Gap-Filling Performance on the Large-Sample Test Set}

We further evaluated RadarFill over 4,384 test times from 7 June to 31 December 2025 and compared it with fill\_data\_3D and 3D U-Net-GAN (Fig. 4). Over 1--8 km, layer-wise RMSE for RadarFill was mainly 6--9 dBZ, while bias was generally negative at approximately $-$1.5 to $-$5 dBZ; CSI at 15 dBZ varied with altitude (Fig. 4a--c). Under the same evaluation conditions, fill\_data\_3D and 3D U-Net-GAN generally showed larger RMSE and more negative bias, with CSI lower than that of RadarFill at most altitude levels; the largest errors were observed for fill\_data\_3D. These large-sample statistics are consistent with the low-level recovery in the representative weather cases, indicating that the recovery of principal echoes was not limited to a small number of examples.

At reflectivity thresholds of 15, 25, 35, and 45 dBZ, RadarFill achieved RMSE values of 7.36, 7.21, 8.45, and 11.56 dBZ, bias values of $-$2.00, $-$2.78, $-$5.04, and $-$8.92 dBZ, and CSI values of 0.839, 0.732, 0.485, and 0.272, respectively (Fig. 4d--f). With increasing threshold, CSI decreased, whereas RMSE and the magnitude of the negative bias generally increased. fill\_data\_3D and 3D U-Net-GAN showed similar threshold-dependent behavior, but both exhibited larger continuous errors and larger absolute bias overall. RadarFill achieved higher CSI than both reference methods at 15, 25, and 35 dBZ, while CSI decreased further for all three methods at 45 dBZ.

Similar behavior was found for composite reflectivity (Fig. 4g--i). At thresholds of 15, 25, 35, and 45 dBZ, RadarFill achieved RMSE values of 3.39, 3.19, 3.43, and 4.87 dBZ, bias values of $-$0.28, $-$0.53, $-$1.17, and $-$2.78 dBZ, and CSI values of 0.923, 0.870, 0.784, and 0.675, respectively. Compared with fill\_data\_3D and 3D U-Net-GAN, RadarFill had lower RMSE and bias closer to zero at all four thresholds. Its CSI was also higher at 15, 25, and 35 dBZ, whereas 3D U-Net-GAN achieved a slightly higher CSI at 45 dBZ. Overall, the 4,384-test-time statistics show that RadarFill can recover the principal moderate-to-strong echo regions with relatively stable performance under extensive low-level missingness, while maintaining lower reflectivity-magnitude errors and smaller systematic biases than the reference methods at most evaluation thresholds.

An inference-only ablation further indicates that imposing available observations at each reverse sampling step reduces reflectivity-magnitude errors compared with conditioning and final observation preservation alone (Supplementary Section S3).

\begin{figure}[H]
\centering
\includegraphics[width=\linewidth,height=0.73\textheight,keepaspectratio]{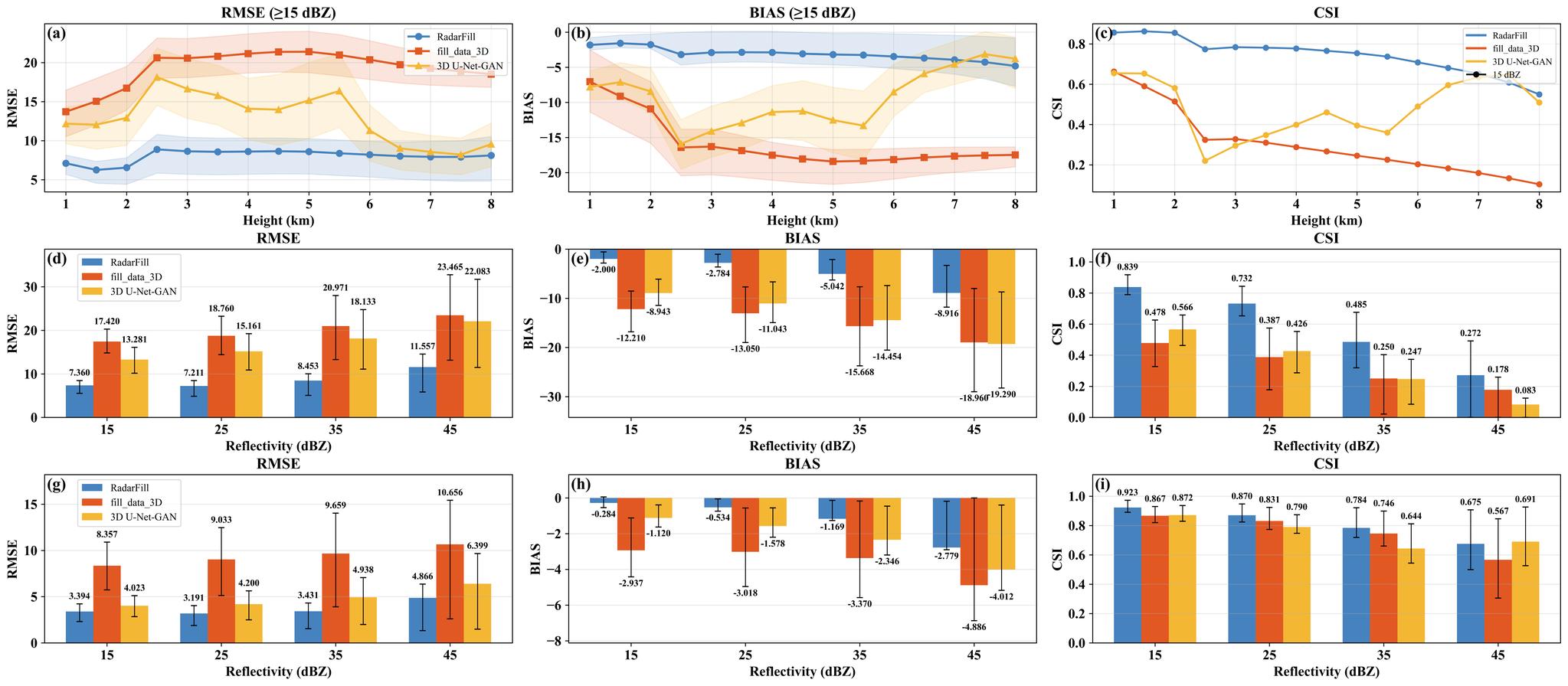}
\caption{Statistics of 3-D radar-reflectivity gap-filling performance on the large-sample test set. RadarFill, the conventional interpolation method fill\_data\_3D, and the learning-based gap-filling reference model 3D U-Net-GAN are evaluated over 4,384 test times from 7 June to 31 December 2025. (a--c) Gap-filling performance at different altitudes over 1--8 km: (a) root-mean-square error (RMSE) for samples with observed reflectivity $\geq$15 dBZ; (b) corresponding mean Bias; and (c) critical success index (CSI) at 15 dBZ. (d--f) Statistics within the 3-D missing region at different reflectivity thresholds: (d) RMSE, (e) Bias and (f) CSI. (g--i) Composite-reflectivity evaluation: (g) RMSE, (h) Bias and (i) CSI. Shading and error bars represent the 25th--75th percentile range (interquartile range, IQR) of the corresponding sample-wise metric distributions.}\label{fig:4}
\end{figure}

To further examine how these statistical differences appear in individual low-level echo structures, a test time from 29 July 2025 was selected for comparison (Fig. 5). At 1.0--2.5 km, RadarFill recovered the locations, morphology, and reflectivity intensity of the principal precipitation echoes in overall agreement with the observation field (Fig. 5a--d,y--ab). In comparison, fill\_data\_3D showed local reflectivity overestimation and echo-shape deviations in parts of the missing region (Fig. 5m--p), whereas 3D U-Net-GAN underestimated moderate-to-strong echo intensity in some areas and produced a smaller recovered echo extent (Fig. 5s--v). The meridional (S--N) and zonal (W--E) cross-sections further show that RadarFill more closely recovered the intensity and vertical extent of the principal echoes in both directions, while both reference methods exhibited more evident intensity biases in some low-level regions (Fig. 5e,f,q,r,w,x,ac,ad). This case illustrates the differences among the three methods in both low-level spatial distribution and vertical structure and is consistent with the large-sample statistics.

\begin{figure}[H]
\centering
\includegraphics[width=\linewidth,height=0.73\textheight,keepaspectratio]{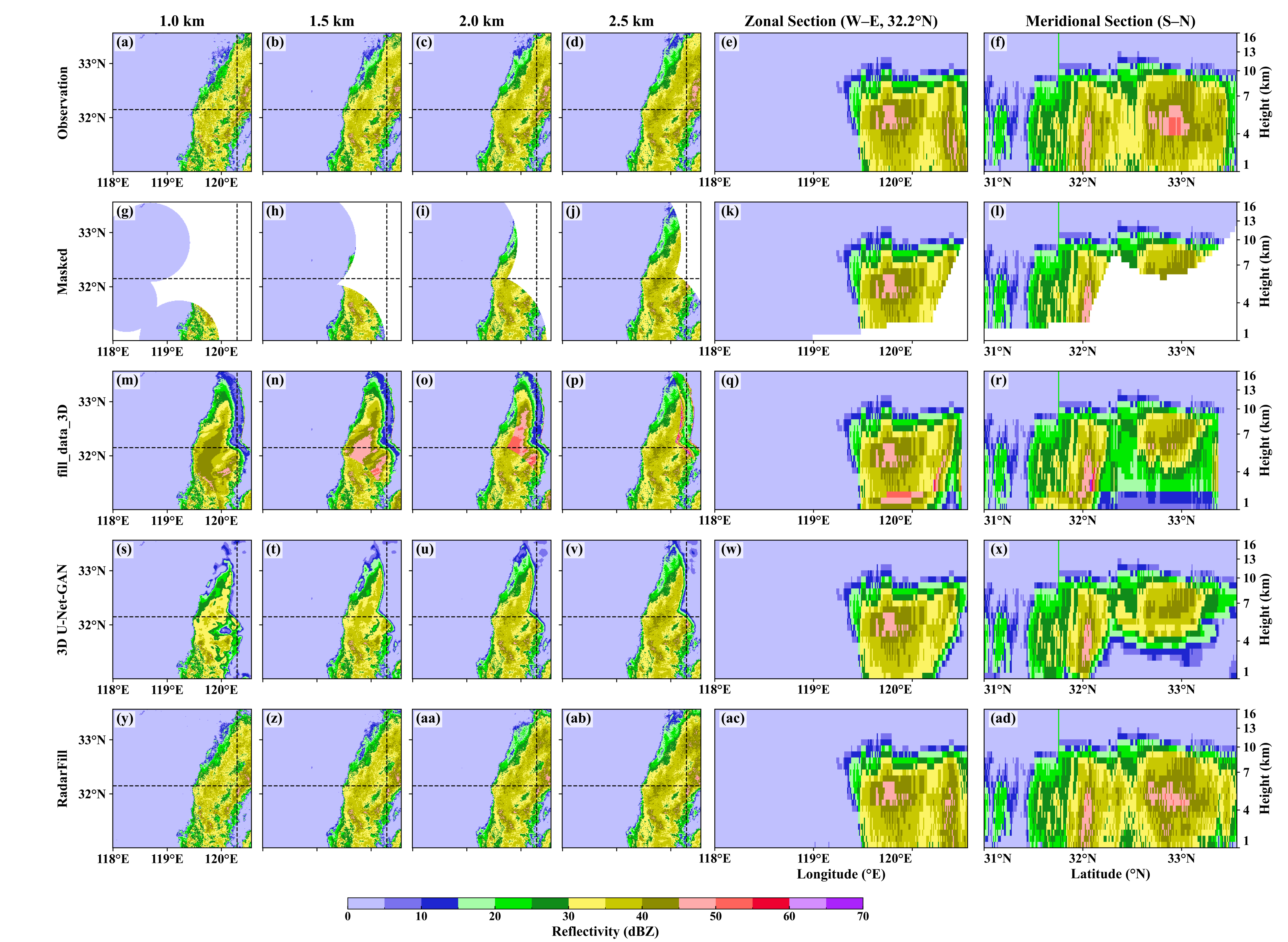}
\caption{Comparison of reflectivity recovery by different gap-filling methods under extensive low-level synthetic missingness. A test time at 23:24 UTC on 29 July 2025 is shown over 118.00$^\circ$E--120.56$^\circ$E and 31.00$^\circ$N--33.56$^\circ$N. RadarFill, fill\_data\_3D and 3D U-Net-GAN are compared under the same structured synthetic missingness. (a--d), (g--j), (m--p), (s--v) and (y--ab) show reflectivity at 1.0, 1.5, 2.0 and 2.5 km for the observation field, synthetically masked input, fill\_data\_3D, 3D U-Net-GAN and RadarFill, respectively. (e), (k), (q), (w) and (ac) are zonal (W--E) vertical cross-sections along 32.15$^\circ$N, and (f), (l), (r), (x) and (ad) are meridional (S--N) vertical cross-sections along 120.29$^\circ$E. Black dashed lines in the constant-altitude panels mark the cross-section locations. All three gap-filling methods preserve the original valid observations and use their respective gap-filled values only at synthetically missing locations.}\label{fig:5}
\end{figure}

\FloatBarrier

\subsection{Cross-Region Validation in Real Blind Zones Using Independent Phased-Array Radar Observations}

To evaluate RadarFill under real blind-zone conditions, the trained model was applied without retraining to Fujian, a region excluded from training, during Typhoon Doksuri on 28 July 2023. The gap-filled CAP fields were evaluated using independent observations from the Minhou S-band dual-polarization phased-array weather radar, which does not contribute to the CAP V3.0 mosaic product. The evaluation used the 90 spatiotemporally matched volume scans and the real-blind-zone evaluation region defined in Section IV-B.

At a representative time, the real CAP blind zone at 1.5 km overlapped part of the precipitation region observed by the phased-array radar. RadarFill recovered the locations and morphology of the principal precipitation echoes within the blind zone in broad agreement with the phased-array observations (Fig. 6b--e). In comparison, fill\_data\_3D recovered the large-scale precipitation coverage but produced a smoother field with pronounced local reflectivity overestimation. The composite-reflectivity fields and corresponding absolute-error distributions likewise showed closer agreement between the principal precipitation structures reconstructed by RadarFill and the phased-array observations (Fig. 6f--j).

Across the 90 matched volume scans, the reflectivity distributions of both gap-filling methods differed to some extent from the phased-array observations. Compared with fill\_data\_3D, the RadarFill distribution more closely followed the phased-array observations over the principal precipitation-echo intensity range, although discrepancies remained at low reflectivity (Fig. 6k). Within the real missing region over 1.0--2.0 km, RadarFill achieved RMSE values of 12.08, 12.22, and 13.40 dBZ at thresholds of 15, 25, and 35 dBZ, respectively, compared with 12.81, 12.38, and 14.44 dBZ for fill\_data\_3D. The corresponding RadarFill CSI values were 0.67, 0.53, and 0.23, with the values at 25 and 35 dBZ exceeding those of fill\_data\_3D (0.39 and 0.09, respectively; Fig. 6l,m). These results indicate useful reflectivity-recovery capability within the principal low-level real blind zones, although the recovery performance varies with altitude.

Taken together, the Fujian experiment shows that RadarFill recovered principal precipitation structures indicated by independent phased-array observations within real CAP blind zones in a region excluded from training. However, CAP and phased-array radar differ in scanning geometry, sampling volume, and data processing, and the present validation covers only one typhoon event. The results should therefore be interpreted as independent observational-consistency evidence in real cross-region blind zones rather than pixel-level ground-truth validation or evidence of general cross-region generalization.

\begin{figure}[H]
\centering
\includegraphics[width=\linewidth,height=0.73\textheight,keepaspectratio]{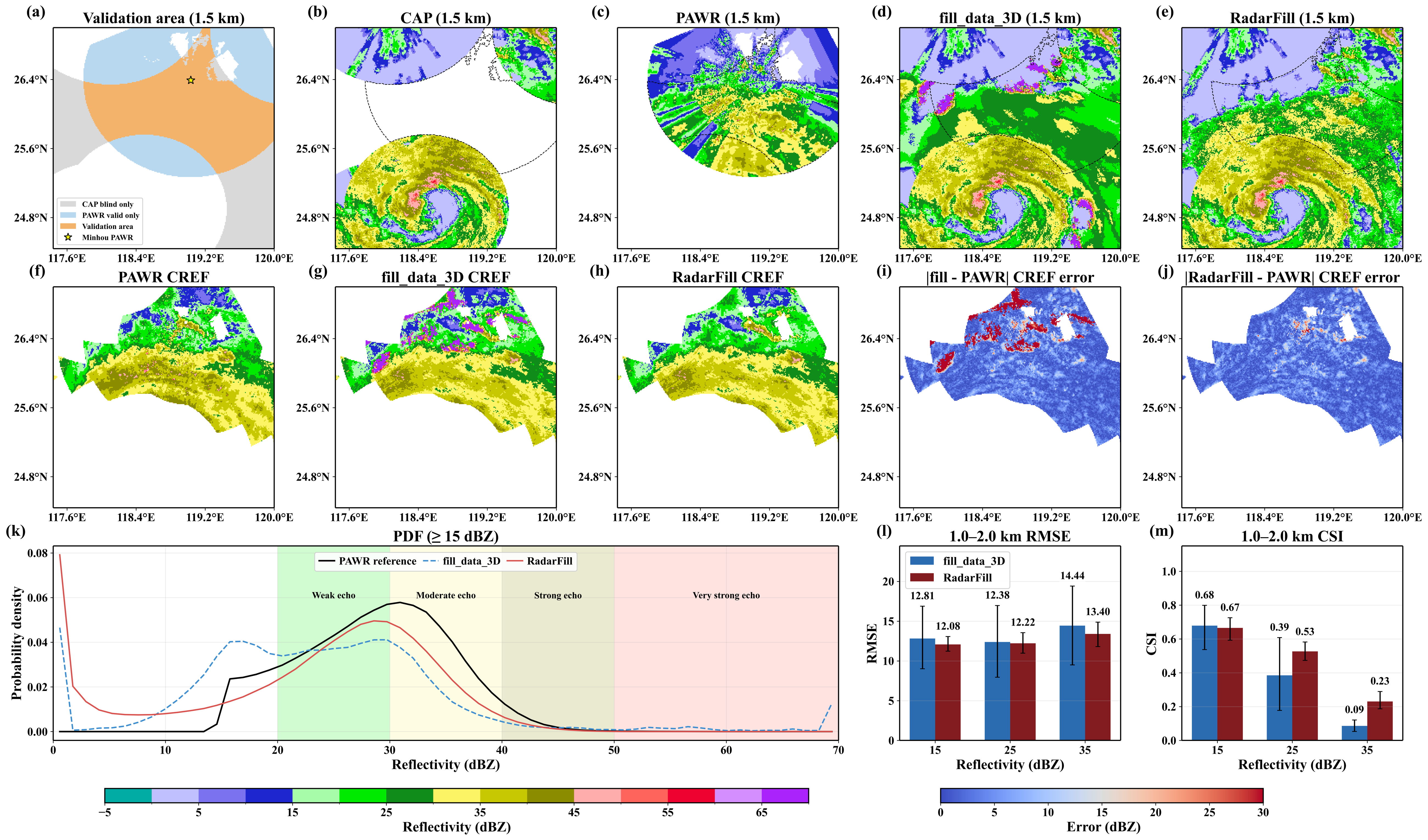}
\caption{Cross-region observational validation of real CAP blind zones using an independent phased-array radar. (a) Validation geometry at 1.5 km altitude, showing the CAP blind zone, valid phased-array observation area and their intersection; the star marks the Minhou phased-array radar. (b--e) Reflectivity at 1.5 km for the original CAP input, independent phased-array observation, fill\_data\_3D gap-filled field and RadarFill gap-filled field, respectively. (f--h) Composite reflectivity (CREF) from the phased-array observation, fill\_data\_3D and RadarFill. (i--j) Absolute composite-reflectivity errors of the two gap-filling methods relative to the phased-array observation. (k) Reflectivity probability density functions over common valid missing samples from 1.0--2.5 km with phased-array reflectivity $\geq$15 dBZ. (l--m) RMSE and CSI within the real missing region over 1.0--2.0 km at thresholds of 15, 25 and 35 dBZ. Bar heights are equally weighted means of scan-wise metrics, and error bars show 95\% confidence intervals estimated using moving-block bootstrap resampling.}\label{fig:6}
\end{figure}

\FloatBarrier

\subsection{Impact of the Gap-Filled Initial Field on Nowcasting}

We next examined whether improvements in the gap-filled 3-D reflectivity initial fields persisted into subsequent nowcasts using the experimental configuration described in Section IV-C. Results are presented primarily for the typhoon event on 30 July 2025, with the precipitation event on 25 August 2025 providing supplementary evaluation. At a representative time during the typhoon event, the CAP composite-reflectivity field showed a broad precipitation system with extensive organized echoes within the study domain (Fig. 7).

RadarFill substantially reduced errors in the 3-D reflectivity initial field, with the largest improvements concentrated at low levels (Fig. 8d--f). Over 1--2.5 km, initial-field RMSE decreased from 8.04 dBZ for Fill to 3.48 dBZ, a reduction of 56.7\%, while bias changed from $-$2.78 to $-$0.42 dBZ, reducing its absolute magnitude by 84.8\%. During the 6--180-min forecasts, CSI at 15, 25, and 35 dBZ remained higher for RadarFill than for Fill and was generally closer to Observation (Fig. 8a--c), with mean relative improvements of 5.95\%, 6.41\%, and 12.41\%, respectively. The RMSE and bias improvements decreased with lead time: at +6 min, RMSE and absolute bias were reduced by 23.3\% and 48.2\%, respectively, whereas the corresponding reductions at +180 min were 3.1\% and 5.9\%. The larger RMSE differences were concentrated at low altitudes and diminished with height, while CSI at all three thresholds remained higher than Fill throughout the 180-min forecast period.

Similar behavior was found for the precipitation event on 25 August 2025 (Supplementary Fig. S1). Over 1--2.5 km, initial-field RMSE decreased from 9.69 to 5.05 dBZ, a reduction of 47.84\%, while bias changed from $-$4.93 to $-$1.22 dBZ, reducing its absolute magnitude by 75.19\%. CSI at 15, 25, and 35 dBZ remained higher than Fill over 6--180 min, with mean relative improvements of 5.22\%, 8.25\%, and 12.85\%, respectively. As in the typhoon event, improvements in RMSE and bias were concentrated mainly at low altitudes and short lead times and generally weakened with increasing lead time. Across the two weather events examined, the improvements introduced by the RadarFill initial fields persisted into subsequent nowcasts, with the clearest influence at low altitudes and short lead times.

\begin{figure}[H]
\centering
\includegraphics[width=\linewidth,height=0.73\textheight,keepaspectratio]{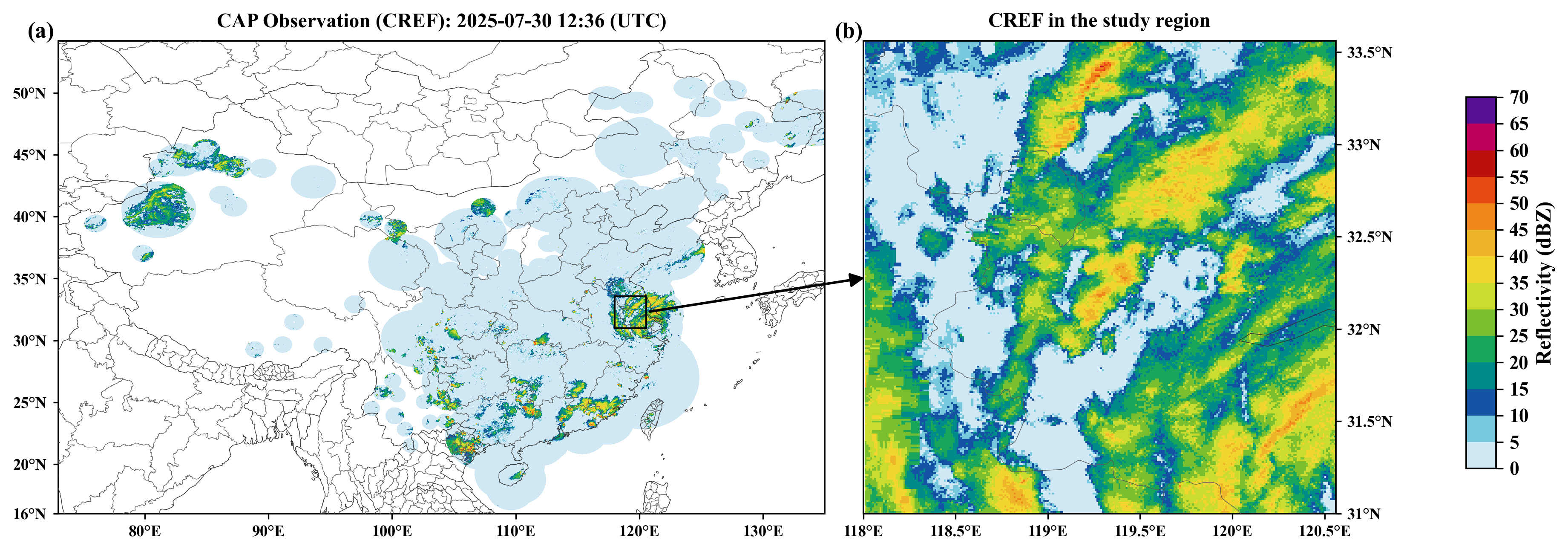}
\caption{(a) CAP composite reflectivity (CREF) over China at 12:36 UTC on 30 July 2025; the black box marks the study domain. (b) Composite-reflectivity distribution within the study domain.}\label{fig:7}
\end{figure}

\begin{figure}[H]
\centering
\includegraphics[width=\linewidth,height=0.73\textheight,keepaspectratio]{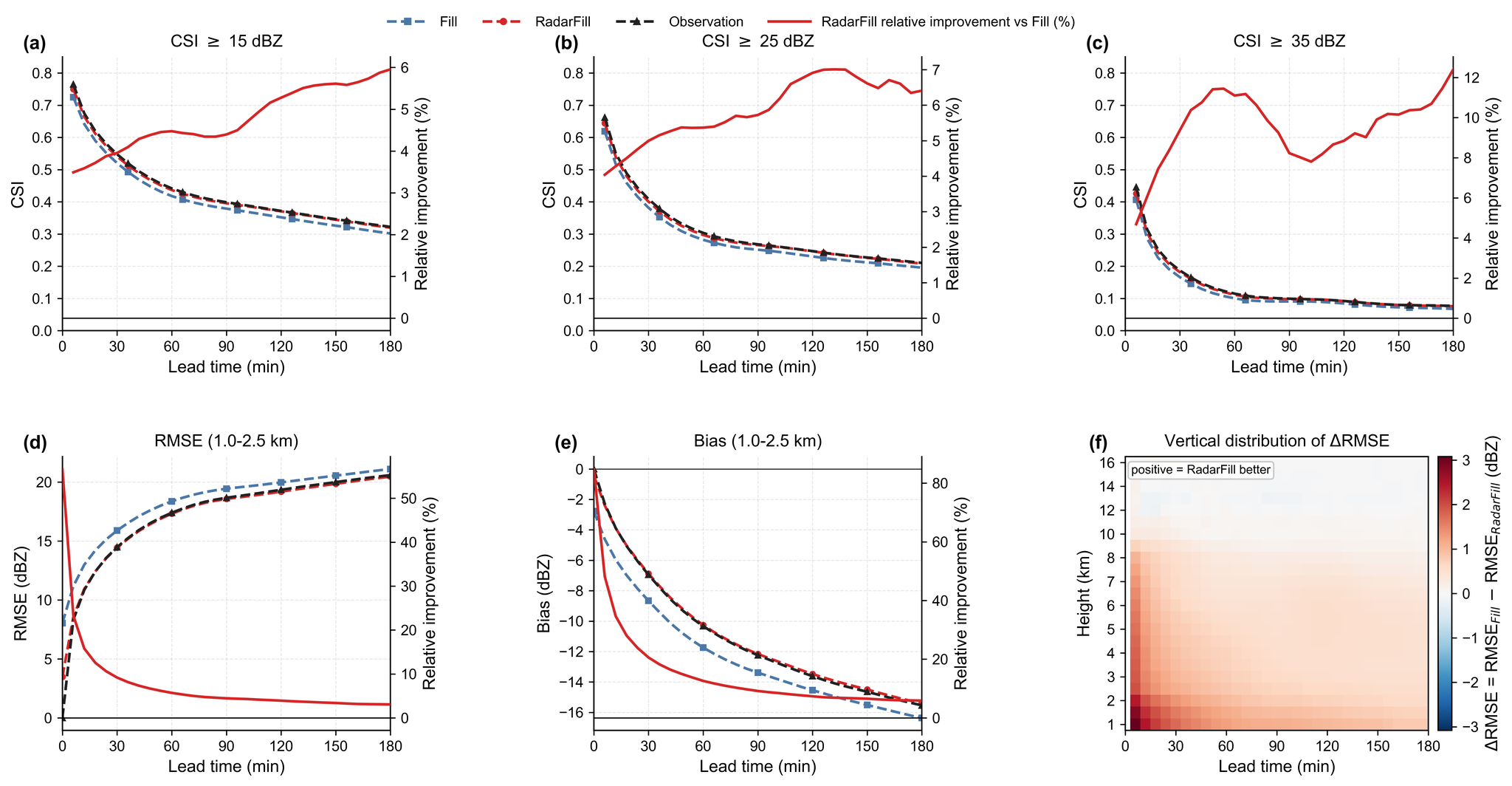}
\caption{Impact of different 3-D reflectivity initial fields on nowcasting performance. The same nowcasting model was initialized with fill\_data\_3D, RadarFill and the complete observation field for 201 forecast cycles during the typhoon event on 30 July 2025, denoted Fill, RadarFill and Observation, respectively. (a--c) Critical success index (CSI) at thresholds of 15, 25 and 35 dBZ as a function of lead time. Dashed lines and the left vertical axes indicate CSI for the three initialization schemes, while the red solid lines and the right vertical axes indicate the relative CSI improvement of RadarFill over Fill. (d--e) Mean RMSE and Bias over 1--2.5 km as functions of lead time. Here 0 min denotes the initial field; dashed lines and the left vertical axes indicate the original errors, while the red solid lines and the right vertical axes indicate the relative improvement of RadarFill over Fill. Positive relative improvement indicates that RadarFill outperforms Fill. (f) Altitude--lead-time distribution of the RMSE difference between RadarFill and Fill, defined as Fill RMSE minus RadarFill RMSE; positive values indicate lower RMSE for RadarFill.}\label{fig:8}
\end{figure}

\FloatBarrier

\section{Discussion}

RadarFill generally recovered the principal precipitation structures and moderate-to-strong echoes under extensive low-level missingness, but its performance varied with reflectivity intensity and spatial scale. In the thunderstorm case, the vertical positions of strong echoes were reproduced relatively well, whereas the horizontal displacement of the reflectivity-weighted centroid increased to 6.96 km at 55 dBZ. In the typhoon case, FSS increased with neighborhood size, while stronger echoes showed lower skill at small spatial scales; the reconstructed reflectivity also exhibited a generally negative bias over 1--8 km. Consistent with these case studies, the large-sample evaluation showed decreasing CSI and increasing reflectivity errors with increasing threshold. These results indicate that RadarFill is more robust in recovering the principal 3-D precipitation structure and moderate-to-strong echoes than in reproducing the intensity and precise local position of extremely strong echoes.

The real-blind-zone experiment provides complementary evidence beyond controlled synthetic missingness. In Fujian, a region excluded from training, RadarFill produced lower reflectivity errors than fill\_data\_3D over the principal 1.0--2.0-km blind zones and recovered precipitation structures broadly consistent with independent phased-array observations. However, CAP and phased-array radar differ in scanning geometry, sampling volume, and data processing, and the validation is limited to one typhoon event. The experiment should therefore be interpreted as independent observational-consistency evidence for real cross-region blind zones rather than strict pixel-level ground-truth validation or evidence of universal cross-region generalization.

\FloatBarrier

\section{Conclusion}

We present RadarFill for filling systematic blind zones in 3-D radar reflectivity mosaics using conditional generation constrained by available radar observations. Controlled evaluations under structured synthetic missingness show that RadarFill can recover the principal precipitation structures and moderate-to-strong echoes under extensive low-level missingness and generally outperforms fill\_data\_3D and 3D U-Net-GAN in reflectivity error and echo-detection skill, although limitations remain for stronger echoes. Independent phased-array observations in Fujian further provide observational-consistency evidence that the learned gap-filling capability may extend to real CAP blind zones in a region excluded from training. In addition, the improvements in the reconstructed initial fields persisted into subsequent Nowcast3D forecasts in the two weather events examined. Overall, these results support observation-constrained conditional generation as a practical approach for improving the completeness of 3-D radar reflectivity mosaics and motivate further evaluation across additional regions, weather systems, and real blind-zone conditions.

\clearpage
\section*{Supplementary Material}
\setcounter{figure}{0}
\setcounter{table}{0}
\renewcommand{\thefigure}{S\arabic{figure}}
\renewcommand{\theHfigure}{supp.\arabic{figure}}
\renewcommand{\thetable}{S\arabic{table}}
\renewcommand{\theHtable}{supp.\arabic{table}}
\FloatBarrier

\subsection*{S1. Vertical Structure of Real Missing Regions in the CAP Mosaic Product}

To characterize the spatial correspondence of real missing regions across altitude in the CAP 3D reflectivity mosaic product, we calculated the vertical containment rate of missing regions between adjacent altitudes over Zhejiang (28.00$^\circ$--30.56$^\circ$N, 119.00$^\circ$--121.56$^\circ$E) and the region surrounding Shanghai (31.00$^\circ$--33.56$^\circ$N, 120.00$^\circ$--122.56$^\circ$E). For adjacent lower and upper levels, the containment rate is defined as

\[R_{\mathrm{inc}}=\frac{|D_{\mathrm{high}}\cap D_{\mathrm{low}}|}{|D_{\mathrm{high}}|}\]

where \(\text{D}_{\text{low}}\) and \(\text{D}_{\text{high}}\) denote the sets of missing pixels at the adjacent lower and upper levels, respectively. The metric gives the fraction of missing pixels at the upper level that also lie within the missing region at the lower level. The containment rate is first calculated independently at each time and then averaged with equal weight over times with valid statistics. The results show a pronounced vertical spatial correspondence of real CAP missingness at the principal low levels: within 1.0--2.5 km, approximately 77.69\%--99.97\% of missing pixels at the higher level also lie within the missing region at the adjacent lower level on average. These statistics are used to define the overall pattern of structured synthetic missingness---broad at low levels and generally contracting with altitude---rather than to fit the real blind zone of any specific region.

\begingroup
\footnotesize
\setlength{\tabcolsep}{3pt}
\begin{longtable}{@{}>{\raggedright\arraybackslash}p{0.0632\linewidth}>{\raggedright\arraybackslash}p{0.0626\linewidth}>{\raggedright\arraybackslash}p{0.1718\linewidth}>{\raggedright\arraybackslash}p{0.1493\linewidth}>{\raggedright\arraybackslash}p{0.2277\linewidth}>{\raggedright\arraybackslash}p{0.2056\linewidth}@{}}
\caption{Vertical containment rates of real missing regions in the CAP mosaic product (altitude in km)}\\

\toprule\noalign{}
\begin{minipage}[b]{\linewidth}\raggedright
\textbf{Lower\\
level}\strut
\end{minipage} & \begin{minipage}[b]{\linewidth}\raggedright
\textbf{Upper\\
level}\strut
\end{minipage} & \begin{minipage}[b]{\linewidth}\raggedright
\textbf{Zhejiang containment rate}
\end{minipage} & \begin{minipage}[b]{\linewidth}\raggedright
\textbf{Zhejiang sample count}
\end{minipage} & \begin{minipage}[b]{\linewidth}\raggedright
\textbf{Shanghai-adjacent containment rate}
\end{minipage} & \begin{minipage}[b]{\linewidth}\raggedright
\textbf{Shanghai-adjacent sample count}
\end{minipage} \\
\midrule\noalign{}
\endhead
\bottomrule\noalign{}
\endlastfoot
0.5 & 1.0 & 99.8839\% & 17534 & 99.7907\% & 17534 \\
1.0 & 1.5 & 99.5383\% & 17534 & 99.9441\% & 17534 \\
1.5 & 2.0 & 77.6905\% & 17534 & 99.9364\% & 17534 \\
2.0 & 2.5 & 89.6715\% & 17534 & 99.9696\% & 17534 \\
2.5 & 3.0 & 75.1218\% & 17534 & 99.9915\% & 17534 \\
3.0 & 3.5 & 59.7481\% & 17534 & 99.9515\% & 17534 \\
3.5 & 4.0 & 81.0209\% & 17534 & 99.8981\% & 17534 \\
4.0 & 4.5 & 82.8461\% & 17534 & 99.6013\% & 17534 \\
4.5 & 5.0 & 84.3779\% & 17534 & 99.1680\% & 17534 \\
5.0 & 5.5 & 84.7800\% & 17534 & 99.4844\% & 17534 \\
5.5 & 6.0 & 85.9147\% & 17534 & 99.6041\% & 17534 \\
6.0 & 6.5 & 79.1773\% & 17534 & 99.5275\% & 17534 \\
6.5 & 7.0 & 79.6875\% & 17534 & 99.6413\% & 17534 \\
7.0 & 7.5 & 89.5259\% & 17534 & 99.7796\% & 17534 \\
7.5 & 8.0 & 52.4206\% & 17534 & 98.9084\% & 17534 \\
8.0 & 9.0 & 53.9870\% & 17534 & 99.0308\% & 17534 \\
9.0 & 10.0 & 90.1322\% & 17534 & 98.9161\% & 17534 \\
10.0 & 11.0 & 78.4187\% & 17534 & 98.9363\% & 17534 \\
11.0 & 12.0 & 67.7181\% & 17534 & 99.0111\% & 17534 \\
12.0 & 13.0 & 54.3334\% & 17534 & 98.1979\% & 17534 \\
13.0 & 14.0 & 13.6520\% & 17534 & 15.1113\% & 17534 \\
14.0 & 15.0 & 12.4833\% & 17534 & 8.4260\% & 17534 \\
15.0 & 16.0 & 10.2889\% & 17534 & 16.9974\% & 17534 \\
\end{longtable}
\endgroup

Note: The Zhejiang region spans 28.00$^\circ$--30.56$^\circ$N and 119.00$^\circ$--121.56$^\circ$E; the region surrounding Shanghai spans 31.00$^\circ$--33.56$^\circ$N and 120.00$^\circ$--122.56$^\circ$E. The containment rate is the fraction of missing pixels at the upper level that also lie within the missing region at the adjacent lower level. Containment rates are first calculated independently at each time and then averaged with equal weight over times with valid statistics. The results show a pronounced vertically nested structure of real low-level missing regions in the CAP mosaic product.

\FloatBarrier

\subsection*{S2. Nowcasting Experiment for the Precipitation Event on 25 August 2025}

To further examine the impact of the improved initial field on subsequent nowcasts, we conducted a supplementary experiment for a precipitation event on 25 August 2025, comprising 201 forecast cycles. The experiment used the same synthetic missingness configuration and pretrained Nowcast3D model as described in Section IV-C of the main text. The Fill, RadarFill and Observation experiments used the fill\_data\_3D gap-filled fields, RadarFill gap-filled fields and observation fields as historical inputs, respectively, with lead times from 6 to 180 min.

For this precipitation event, RadarFill reduced errors in the 3D reflectivity initial field, with the largest improvement concentrated at low levels (Supplementary Fig. S1d--f). Over 1.0--2.5 km, initial-field RMSE decreased by 47.84\%, from 9.69 dBZ for Fill to 5.05 dBZ, while Bias improved from $-$4.93 to $-$1.22 dBZ, reducing the bias magnitude by 75.19\%. During the forecast, CSI at 15, 25 and 35 dBZ remained higher for RadarFill than for Fill over 6--180 min and was generally closer to Observation (Supplementary Fig. S1a--c). Mean absolute CSI increments at the three thresholds were 0.0129, 0.0127 and 0.0076, respectively, with corresponding mean relative improvements of 5.22\%, 8.25\% and 12.85\%. As lead time increased, the absolute CSI increments generally decreased, whereas the relative improvements increased, particularly at 35 dBZ.

Forecast error differences resulting from the improved initial field gradually weakened with lead time (Supplementary Fig. S1d--f). At +6 min, RadarFill reduced RMSE by 22.93\% relative to Fill and reduced the absolute Bias by 49.40\%; by +180 min, these improvements had decreased to 2.12\% and 5.57\%, respectively. In the vertical, the larger RMSE differences were concentrated at low levels and short lead times and generally diminished with increasing altitude and lead time. Notably, although the relative CSI improvement at 35 dBZ increased with lead time, the absolute CSI increment decreased from 0.0133 at +6 min to 0.0049 at +180 min, indicating that an increase in relative improvement at longer lead times does not necessarily imply a larger absolute forecast gain. The results for this precipitation event further show that improvements in reflectivity errors introduced by initial-field gap filling can persist into subsequent nowcasts, although their influence gradually weakens with increasing lead time.

\begin{figure}[H]
\centering
\includegraphics[width=\linewidth,height=0.75\textheight,keepaspectratio]{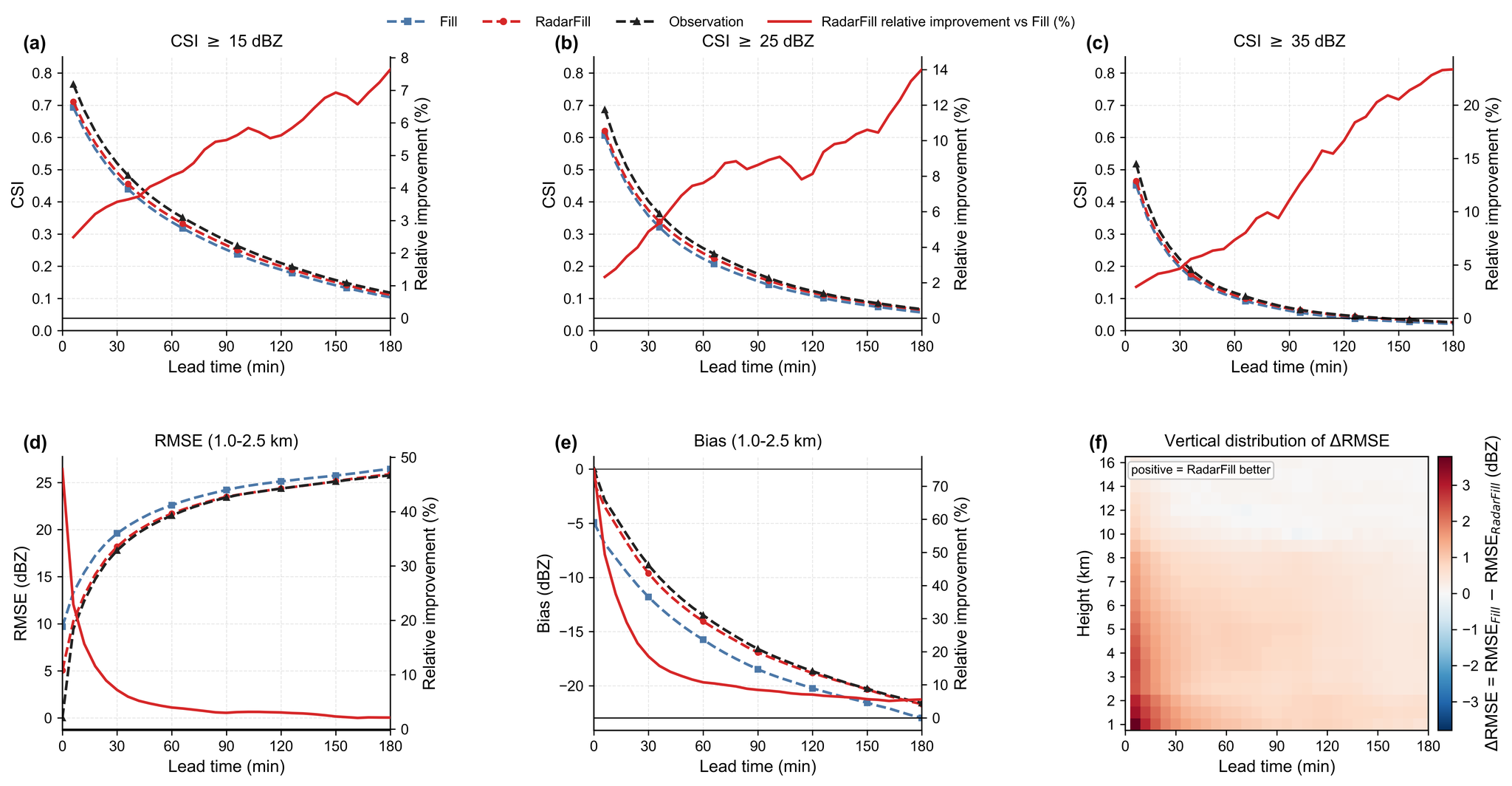}
\captionsetup{name={Supplementary Fig.}}
\caption{Impact of different 3D reflectivity initial fields on nowcasting performance for the precipitation event on 25 August 2025. The same nowcasting model was initialized with fill\_data\_3D, RadarFill and the complete observation field for 201 forecast cycles, denoted Fill, RadarFill and Observation, respectively. (a--c) Critical success index (CSI) at thresholds of 15, 25 and 35 dBZ as a function of lead time. Dashed lines and the left vertical axes indicate CSI for the three initialization schemes, while the red solid lines and the right vertical axes indicate the relative CSI improvement of RadarFill over Fill. (d--e) Mean RMSE and Bias over 1--2.5 km as functions of lead time. Here 0 min denotes the initial field; dashed lines and the left vertical axes indicate the original errors, while the red solid lines and the right vertical axes indicate the relative improvement of RadarFill over Fill. Positive relative improvement indicates that RadarFill outperforms Fill. (f) Altitude--lead-time distribution of the RMSE difference between RadarFill and Fill, defined as Fill RMSE minus RadarFill RMSE; positive values indicate lower RMSE for RadarFill.}\label{fig:S1}
\end{figure}

\FloatBarrier

\subsection*{S3. Ablation of the Reverse-Step Observation Constraint}

To assess the contribution of the reverse-step observation constraint, we compared the standard RadarFill sampler with a variant that omitted the stepwise replacement of observed locations while retaining the same trained model, conditioning information, sampling settings, and final preservation of valid observations. The ablation experiment included all 3,979 valid held-out test samples from 7 June to 6 November 2025, with identical synthetic missingness masks used for both samplers at each test time.

At the 15-dBZ threshold, the standard sampler yielded lower mean RMSE and less negative Bias at all 23 altitude levels (Supplementary Table S2). Similar improvements were found within the 3-D masked region and for composite reflectivity across the evaluated thresholds (Supplementary Table S3). For the 3-D masked region, mean RMSE at 15 and 25 dBZ decreased from 7.98 and 7.93 dBZ to 7.29 and 7.06 dBZ, respectively, while the corresponding Bias improved from $-$2.77 and $-$3.31 dBZ to $-$1.92 and $-$2.60 dBZ. These results indicate that the reverse-step observation constraint helps reduce reflectivity-magnitude errors and systematic underestimation beyond conditioning and final observation preservation alone.

\begingroup
\footnotesize
\renewcommand{\arraystretch}{1.0}
\setlength{\tabcolsep}{3pt}
\begin{longtable}{@{}>{\raggedright\arraybackslash}p{0.147\linewidth}>{\raggedright\arraybackslash}p{0.1486\linewidth}>{\raggedright\arraybackslash}p{0.212\linewidth}>{\raggedright\arraybackslash}p{0.1238\linewidth}>{\raggedright\arraybackslash}p{0.1618\linewidth}>{\raggedright\arraybackslash}p{0.0867\linewidth}@{}}
\caption{LAYER-WISE RMSE AND BIAS WITH AND WITHOUT THE REVERSE-STEP OBSERVATION CONSTRAINT AT THE 15-dBZ THRESHOLD}\\

\toprule\noalign{}
\begin{minipage}[b]{\linewidth}\raggedright
\textbf{Altitude (km)}
\end{minipage} & \begin{minipage}[b]{\linewidth}\raggedright
\textbf{Full RMSE}
\end{minipage} & \begin{minipage}[b]{\linewidth}\raggedright
\textbf{Final-only RMSE}
\end{minipage} & \begin{minipage}[b]{\linewidth}\raggedright
\textbf{Full Bias}
\end{minipage} & \begin{minipage}[b]{\linewidth}\raggedright
\textbf{Final-only Bias}
\end{minipage} & \begin{minipage}[b]{\linewidth}\raggedright
\textbf{N}
\end{minipage} \\
\midrule\noalign{}
\endhead
\bottomrule\noalign{}
\endlastfoot
1.0 & 6.76 & 7.05 & $-$1.53 & $-$1.79 & 1956 \\
1.5 & 6.28 & 6.72 & $-$1.60 & $-$2.02 & 1953 \\
2.0 & 6.59 & 7.18 & $-$1.78 & $-$2.41 & 1929 \\
2.5 & 8.86 & 9.77 & $-$3.03 & $-$4.38 & 1866 \\
3.0 & 8.57 & 9.50 & $-$2.75 & $-$4.08 & 1866 \\
3.5 & 8.48 & 9.44 & $-$2.67 & $-$4.05 & 1860 \\
4.0 & 8.47 & 9.47 & $-$2.53 & $-$3.98 & 1846 \\
4.5 & 8.52 & 9.55 & $-$2.65 & $-$4.19 & 1838 \\
5.0 & 8.57 & 9.63 & $-$2.88 & $-$4.47 & 1829 \\
5.5 & 8.36 & 9.42 & $-$2.88 & $-$4.46 & 1819 \\
6.0 & 8.10 & 9.12 & $-$2.97 & $-$4.48 & 1809 \\
6.5 & 7.82 & 8.83 & $-$3.11 & $-$4.54 & 1788 \\
7.0 & 7.68 & 8.67 & $-$3.35 & $-$4.70 & 1766 \\
7.5 & 7.62 & 8.60 & $-$3.64 & $-$4.94 & 1720 \\
8.0 & 7.72 & 8.68 & $-$4.12 & $-$5.35 & 1657 \\
9.0 & 7.67 & 8.52 & $-$5.16 & $-$6.12 & 1488 \\
10.0 & 8.48 & 9.13 & $-$6.91 & $-$7.56 & 1160 \\
11.0 & 9.80 & 10.35 & $-$8.67 & $-$9.22 & 683 \\
12.0 & 11.67 & 12.16 & $-$10.81 & $-$11.31 & 409 \\
13.0 & 13.11 & 13.59 & $-$12.29 & $-$12.77 & 257 \\
14.0 & 14.82 & 15.28 & $-$14.16 & $-$14.63 & 174 \\
15.0 & 17.22 & 17.51 & $-$16.81 & $-$17.13 & 120 \\
16.0 & 19.18 & 19.35 & $-$18.98 & $-$19.15 & 71 \\
\end{longtable}
\endgroup

Note: RMSE and Bias are in dBZ. Full denotes the standard RadarFill sampler with reverse-step observation constraints; Final-only omits these constraints while retaining conditioning and final observation preservation. Metrics are calculated within the synthetically masked region where reference reflectivity is $\geq$15 dBZ. Values are means of valid time-wise metrics, and N denotes the number of valid test times at each altitude.

\begingroup
\footnotesize
\renewcommand{\arraystretch}{1.0}
\setlength{\tabcolsep}{3pt}
\begin{longtable}{@{}>{\raggedright\arraybackslash}p{0.1333\linewidth}>{\raggedright\arraybackslash}p{0.1199\linewidth}>{\raggedright\arraybackslash}p{0.1541\linewidth}>{\raggedright\arraybackslash}p{0.1402\linewidth}>{\raggedright\arraybackslash}p{0.1244\linewidth}>{\raggedright\arraybackslash}p{0.1265\linewidth}>{\raggedright\arraybackslash}p{0.0815\linewidth}@{}}
\caption{RMSE AND BIAS WITH AND WITHOUT THE REVERSE-STEP OBSERVATION CONSTRAINT AT DIFFERENT REFLECTIVITY THRESHOLDS}\\

\toprule\noalign{}
\begin{minipage}[b]{\linewidth}\raggedright
\textbf{Region}
\end{minipage} & \begin{minipage}[b]{\linewidth}\raggedright
\textbf{Threshold (dBZ)}
\end{minipage} & \begin{minipage}[b]{\linewidth}\raggedright
\textbf{Full RMSE}
\end{minipage} & \begin{minipage}[b]{\linewidth}\raggedright
\textbf{Final-only RMSE}
\end{minipage} & \begin{minipage}[b]{\linewidth}\raggedright
\textbf{Full Bias}
\end{minipage} & \begin{minipage}[b]{\linewidth}\raggedright
\textbf{Final-only Bias}
\end{minipage} & \begin{minipage}[b]{\linewidth}\raggedright
\textbf{N}
\end{minipage} \\
\midrule\noalign{}
\endhead
\bottomrule\noalign{}
\endlastfoot
3-D masked region & 15 & 7.29 & 7.98 & $-$1.92 & $-$2.77 & 1956 \\
& 25 & 7.06 & 7.93 & $-$2.60 & $-$3.31 & 1956 \\
& 35 & 7.91 & 8.81 & $-$4.41 & $-$5.02 & 1948 \\
& 45 & 11.07 & 11.80 & $-$8.50 & $-$9.01 & 1763 \\
& 55 & 14.15 & 14.88 & $-$12.75 & $-$13.32 & 492 \\
CREF & 15 & 3.31 & 3.60 & $-$0.30 & $-$0.55 & 1956 \\
& 25 & 3.12 & 3.55 & $-$0.48 & $-$0.71 & 1956 \\
& 35 & 3.20 & 3.68 & $-$0.85 & $-$1.06 & 1956 \\
& 45 & 4.59 & 5.02 & $-$2.35 & $-$2.57 & 1899 \\
& 55 & 6.01 & 6.39 & $-$4.87 & $-$5.12 & 631 \\
\end{longtable}
\endgroup

{\noindent\footnotesize Note: RMSE and Bias are in dBZ. For each threshold, metrics are calculated at synthetically masked locations where the reference reflectivity reaches the corresponding threshold. The 3-D statistics include all masked voxels across the 23 altitude levels. CREF is the maximum reflectivity over the vertical levels and is evaluated at horizontal locations masked at 1.0 km, consistent with the main-text evaluation procedure. Values are means over valid test times; N is the valid-time count.\par}


\clearpage
\begin{thebibliography}{99}
\bibitem{ref1} S. Ravuri et al., ``Skilful precipitation nowcasting using deep generative models of radar,'' Nature, vol. 597, pp. 672--677, 2021, doi: 10.1038/s41586-021-03854-z.

\bibitem{ref2} J. Sun et al., ``Use of NWP for nowcasting convective precipitation: Recent progress and challenges,'' Bull. Amer. Meteorol. Soc., vol. 95, pp. 409--426, 2014, doi: 10.1175/BAMS-D-11-00263.1.

\bibitem{ref3} Y.-S. Tang et al., ``A localized quantitative precipitation estimation for S-band polarimetric radar in Taiwan,'' J. Hydrometeorol., vol. 25, pp. 1697--1712, 2024, doi: 10.1175/JHM-D-23-0205.1.

\bibitem{ref4} M. Simpson et al., ``A supplemental high-resolution weather radar network for the conterminous United States,'' Bull. Amer. Meteorol. Soc., vol. 107, pp. E1430--E1443, 2026, doi: 10.1175/BAMS-D-24-0168.1.

\bibitem{ref5} J. Sun, ``Convective-scale assimilation of radar data: Progress and challenges,'' Q. J. R. Meteorol. Soc., vol. 131, pp. 3439--3463, 2005, doi: 10.1256/qj.05.149.

\bibitem{ref6} N. Gustafsson et al., ``Survey of data assimilation methods for convective-scale numerical weather prediction at operational centres,'' Q. J. R. Meteorol. Soc., vol. 144, pp. 1218--1256, 2018, doi: 10.1002/qj.3179.

\bibitem{ref7} R. A. Maddox, J. Zhang, J. J. Gourley, and K. W. Howard, ``Weather radar coverage over the contiguous United States,'' Weather Forecast., vol. 17, pp. 927--934, 2002, doi: 10.1175/1520-0434(2002)017\textless0927:WRCOTC\textgreater2.0.CO;2.

\bibitem{ref8} National Research Council, Observing Weather and Climate from the Ground Up: A Nationwide Network of Networks. Washington, DC, USA: National Academies Press, 2009, doi: 10.17226/12540.

\bibitem{ref9} S. D. Wang, C. Pei, Z. M. Guo, and N. Shao, ``Evaluations on Chinese next-generation radar coverage and terrain blockage based on SRTM data,'' Climate Environ. Res., vol. 16, pp. 459--468, 2011, in Chinese, doi: 10.3878/j.issn.1006-9585.2011.04.06.

\bibitem{ref10} Y. Liu, C. Li, L. Zhang, H. Wen, and S. Wang, ``Statistical analysis of terrain blockage impacts on the CINRAD network based on DEM data,'' Acta Meteorol. Sinica, vol. 78, pp. 705--720, 2020, in Chinese, doi: 10.11676/qxxb2020.037.

\bibitem{ref11} A. Geiss and J. C. Hardin, ``Inpainting radar missing data regions with deep learning,'' Atmos. Meas. Tech., vol. 14, pp. 7729--7747, 2021, doi: 10.5194/amt-14-7729-2021.

\bibitem{ref12} S. Tan, H. Chen, and W.-C. Lee, ``Deep learning for weather radar beam blockage correction: Potential impacts on QPE over complex terrain in Northern California,'' Artif. Intell. Earth Syst., vol. 4, Art. no. e240111, 2025, doi: 10.1175/AIES-D-24-0111.1.

\bibitem{ref13} A. W. Seed and G. G. S. Pegram, ``Using Kriging to infill gaps in radar data due to ground clutter in real-time,'' in Proc. 5th Int. Symp. Hydrological Applications of Weather Radar, 2001, pp. 73--78.

\bibitem{ref14} P. Zhang, D. Zrnić, and A. Ryzhkov, ``Partial beam blockage correction using polarimetric radar measurements,'' J. Atmos. Oceanic Technol., vol. 30, pp. 861--872, 2013, doi: 10.1175/JTECH-D-12-00075.1.

\bibitem{ref15} A. Gong, H. Chen, and G. Ni, ``Improving the completion of weather radar missing data with deep learning,'' Remote Sens., vol. 15, Art. no. 4568, 2023, doi: 10.3390/rs15184568.

\bibitem{ref16} H. Wang et al., ``A CBAM-integrated deep convolutional neural network for weather radar echo correction to overcome terrain occlusion challenge,'' Adv. Space Res., vol. 77, pp. 7610--7633, 2026, doi: 10.1016/j.asr.2026.01.098.

\bibitem{ref17} X. Yin, Z. Hu, J. Zheng, B. Li, and Y. Zuo, ``Study on radar echo-filling in an occlusion area by a deep learning algorithm,'' Remote Sens., vol. 13, Art. no. 1779, 2021, doi: 10.3390/rs13091779.

\bibitem{ref18} S. Tan and H. Chen, ``A conditional generative adversarial network for weather radar beam blockage correction,'' IEEE Trans. Geosci. Remote Sens., vol. 61, Art. no. 4103014, 2023, doi: 10.1109/TGRS.2023.3286181.

\bibitem{ref19} W. Zhang, X. Zhang, Z. Jin, Y. Wen, and J. Liu, ``Inpainting radar missing data via denoising iterative restoration,'' IEEE J. Sel. Topics Appl. Earth Observ. Remote Sens., vol. 17, pp. 10715--10725, 2024, doi: 10.1109/JSTARS.2024.3400407.

\bibitem{ref20} W. Zhang, X. Zhang, J. Dong, X. Song, and R. Pang, ``CIDM: A comprehensive inpainting diffusion model for missing weather radar data with knowledge guidance,'' ISPRS J. Photogramm. Remote Sens., vol. 221, pp. 299--309, 2025, doi: 10.1016/j.isprsjprs.2025.02.001.

\bibitem{ref21} A. O. Zeghina, V. Enescu, N. Viltard, L. Barthes, and C. Mallet, ``Guided conditional diffusion for correction of weather radar beam blockage artefacts,'' in Foundations of Intelligent Systems: 28th International Symposium, ISMIS 2026. Cham, Switzerland: Springer, 2026, pp. 60--69, doi: 10.1007/978-3-032-32643-0\_6.

\bibitem{ref22} Z. Ding et al., ``Radar observation gap-filling technology enhanced by satellite imager measurements,'' Remote Sens., vol. 18, Art. no. 1205, 2026, doi: 10.3390/rs18081205.

\bibitem{ref23} X. Wang, Y. Liu, J. Du, Y. Qin, Z. Xiang, and Y. Shi, ``Rad3DGFM: A deep learning network for completing data gaps in 3D radar reflectivity mosaics,'' Adv. Atmos. Sci., vol. 43, no. 11, pp. 2518--2535, 2026, doi: 10.1007/s00376-026-5840-z.

\bibitem{ref24} M. Arulraj, V. Petkovic, R. R. Ferraro, and H. Meng, ``Precipitation vertical structure characterization: A feature-based approach,'' J. Hydrometeorol., vol. 24, pp. 2281--2297, 2023, doi: 10.1175/JHM-D-23-0034.1.

\bibitem{ref25} H. Chen et al., ``Nowcast3D: Reliable precipitation nowcasting via gray-box learning,'' arXiv:2511.04659, 2025, doi: 10.48550/arXiv.2511.04659.

\bibitem{ref26} J. Ho, A. Jain, and P. Abbeel, ``Denoising diffusion probabilistic models,'' in Advances in Neural Information Processing Systems, vol. 33, 2020, pp. 6840--6851.

\bibitem{ref27} I. Loshchilov and F. Hutter, ``Decoupled weight decay regularization,'' in Proc. Int. Conf. Learn. Representations, 2019.

\bibitem{ref28} J. Song, C. Meng, and S. Ermon, ``Denoising diffusion implicit models,'' in Proc. Int. Conf. Learn. Representations, 2021.

\bibitem{ref29} N. M. Roberts and H. W. Lean, ``Scale-selective verification of rainfall accumulations from high-resolution forecasts of convective events,'' Mon. Weather Rev., vol. 136, pp. 78--97, 2008, doi: 10.1175/2007MWR2123.1.


\end{thebibliography}
\end{document}